\documentclass[12pt,a4paper]{article}
\pdfoutput=1

\usepackage{parskip}

\usepackage[utf8]{inputenc}
\usepackage[T1]{fontenc}

\usepackage{multirow}
\usepackage{amsmath}
\usepackage{color}
\usepackage{amsfonts}
\usepackage{amssymb}
\usepackage{graphicx}
\usepackage{geometry}
\usepackage{amssymb,epsfig,subfigure}
\usepackage{hyperref}
\usepackage{url}
\usepackage{comment}
\usepackage[font=footnotesize]{caption}
\usepackage{slashed}

\makeatletter
\renewcommand\section{\@startsection {section}{1}{\z@}%
                                 {-3.5ex \@plus -1ex \@minus -.2ex}
                                   {2.3ex \@plus.2ex}%
                                   {\normalfont\large\bfseries}}
\renewcommand\subsection{\@startsection{subsection}{2}{\z@}%
                                   {-3.25ex\@plus -1ex \@minus -.2ex}%
                                     {1.5ex \@plus .2ex}%
                                     {\normalfont\bfseries}}
\renewcommand\subsubsection{\@startsection{subsubsection}{3}{\z@}%
                                   {-3.25ex\@plus -1ex \@minus -.2ex}%
                                     {1.5ex \@plus .2ex}%
                                     {\normalfont\itshape}}
\makeatother

\def\pplogo{\vbox{\kern-\headheight\kern -29pt
\halign{##&##\hfil\cr&{\ppnumber}\cr\rule{0pt}{2.5ex}&\ppdate\cr}}}
\makeatletter
\def\ps@firstpage{\ps@empty \def\@oddhead{\hss\pplogo}%
  \let\@evenhead\@oddhead 
}
\def\maketitle{\par
 \begingroup
 \def\thefootnote{\fnsymbol{footnote}}
 \def\@makefnmark{\hbox{$^{\@thefnmark}$\hss}}
 \if@twocolumn
 \twocolumn[\@maketitle]
 \else \newpage
 \global\@topnum\z@ \@maketitle \fi\thispagestyle{firstpage}\@thanks
 \endgroup
 \setcounter{footnote}{0}
 \let\maketitle\relax
 \let\@maketitle\relax
 \gdef\@thanks{}\gdef\@author{}\gdef\@title{}\let\thanks\relax}
\makeatother

\numberwithin{equation}{section}

\newcommand\eea{\end{eqnarray}}
\newcommand\bea{\begin{eqnarray}}

\def\beq{\begin{equation}}
\def\eeq{\end{equation}}

\newcommand{\be}{\begin{equation}}
\newcommand{\ee}{\end{equation}}
\newcommand{\ba}{\begin{align}}
\newcommand{\ea}{\end{align}}
\newcommand{\bg}{\begin{gather}}
\newcommand{\eg}{\end{gather}}
\newcommand{\bseq}{\begin{subequations}}
\newcommand{\eseq}{\end{subequations}}

\renewcommand{\tanh}{\mathop{\rm th}\nolimits}

\renewcommand{\t}{\tilde}

\newcommand{\tr}{{\rm tr}}
\newcommand{\mc}{\mathcal}

\newcommand{\coment}[1]{}

\begin{document}
\setcounter{page}0
\def\ppnumber{\vbox{\baselineskip14pt
}}
\def\ppdate{
} \date{}

\author{Germ\'an Mato$^{1,2}$, Facundo Rigatuso$^{1,2}$, Gonzalo Torroba$^{1, 2}$\\
[7mm] \\
{\normalsize \it $^1$ Instituto Balseiro, UNCuyo and CNEA}\\
{\normalsize \it $^2$Centro At\'omico Bariloche and CONICET}\\
{\normalsize \it S.C. de Bariloche, R\'io Negro, R8402AGP, Argentina}\\
}

\bigskip
\title{\bf  Statistics of correlations in nonlinear recurrent neural networks
\vskip 0.5cm}
\maketitle

\begin{abstract}
The statistics of correlations are central quantities characterizing the collective dynamics of recurrent neural networks. We derive exact expressions for the statistics of correlations of nonlinear recurrent networks in the limit of a large number $N$ of neurons, including systematic $1/N$ corrections,  in the regime of Gaussian quenched disorder. Our approach uses a path-integral representation of the network’s stochastic dynamics, which reduces the description to a few collective variables and enables efficient computation. This generalizes previous results on linear networks to include a wide family of  nonlinear activation functions, which enter as interaction terms in the path integral. These interactions can resolve the instability of the linear theory and yield a strictly positive participation dimension. We present explicit results for power-law activations, revealing scaling behavior controlled by the network coupling. In addition, we introduce a class of activation functions based on Padé approximants and provide analytic predictions for their correlation statistics. Numerical simulations confirm our theoretical results with excellent agreement.  We also compare with previous works that have studied the complementary case with annealed disorder, and based on this we propose a conjectural self-consistent equation for the more general case of colored noise.
\end{abstract}
\bigskip


\newpage
\tableofcontents

\vskip 1cm

\section{Introduction}\label{sec:intro}

Correlations of neural activity are one of the main tools we have to study the structure and function of the nervous system. They provide crucial information by measuring the statistical interdependencies between the activity of different neurons or neural regions over time \cite{cohen2011measuring}. The interpretation of the correlations is not easy  because they are not only controlled by direct interactions between the neurons but also by the global dynamical state of the system. For instance, theoretical analysis reveals that if the network is in an asynchronous state the cross-correlations are  smaller than the autocorrelations by a factor $1/N$, $N$ being the network size \cite{ginzburg1994theory}. This regime has been found to be present in a wide variety of systems \cite{abbott1993asynchronous,amit1997model,brunel1999fast,hertz2010cross} under quite general conditions.

Small cross-correlations are indeed found in cortical recordings both in spontaneous and active states \cite{bair2001correlated,constantinidis2002correlated}.
Even if the cross-correlations are small it has been recognized that they can be important for the coding and information transmission properties of networks \cite{hertz2010cross}. For instance, in a study where correlated firing in MT (Middle Temporal area) was measured, it was found that spike counts from adjacent neurons were noisy and only weakly correlated but that even this small amount of correlated noise placed substantial limits on the benefits of signal averaging across a pool \cite{zohary1994correlated}. It has been also proposed that weak correlations can have a crucial effect in controlling neural variability. In \cite{pattadkal2024synchrony} it is shown that physiological levels of synchrony are sufficient to generate irregular responses found in vivo.

Another important effect of small cross-correlations can be found in the control of the effective dimension of the neural dynamics. Dimensionality is a crucial aspect of the network functionality. As has been observed in \cite{fusi2016neurons}  high-dimensional representations allow a simple linear readout to generate a large number of different potential responses. In contrast, neural representations based on highly specialized neurons are low dimensional and they preclude a linear readout but display better generalization capabilities. One of the most common tools for quantifying the dimension is the {\it participation dimension} \cite{gao2017theory}. This quantity, which is defined in terms of the eigenvalues of the covariance matrix, is a natural continuous measure of dimensionality. It can  also be expressed in terms of the first and second moments of the diagonal and non diagonal terms of the covariance matrix. It can be easily seen that even if the cross-correlations are order $1/N$ they have a crucial effect on the participation dimension in the large $N$ limit because there are order $N$ inputs for each neuron. In fact the leading contribution comes from the relative dispersions of cross-covariances across neurons \cite{dahmen2020strong}.

This implies that a theoretical understanding of the cross-correlations is necessary even if they scale like $1/N$ in the large $N$ limit. The calculation of the relevant moments of the covariance matrix was performed in \cite{doi:10.1073/pnas.1818972116} for random connectivity matrices and linear dynamics. In this work the moments  were evaluated using a  path-integral representation up to order $1/N$, allowing to evaluate the participation dimension. The same final result was also derived in \cite{hu2022spectrum} using techniques of random matrices.
One limitation of these results is that in the linear regime the dynamics becomes unstable when the variance of the weights is too strong. This instability can be suppressed in more realistic systems if sub-linear input -- output transfer functions are introduced.

In this paper we present a theoretical analysis of recurrent neural networks with non-linear transfer functions and Gaussian noise, using path-integral methods. This will lead to a new and conceptually simple representation of the neural correlations in terms of a few collective variables. It will allow us to compute the moments of the covariance matrix including the corrections of order $1/N$ necessary to evaluate the participation dimension. This approach can also be extended to compute all higher order neural correlators, as well as to systematically include subleading $1/N$ corrections. Nonlinear recurrent networks have also been studied with other methods, such as dynamical mean field theory (DMFT) (see for instance \cite{clark2023dimension} and references therein), and random matrix theory \cite{zhang2011nonlinear}.  In fact, as we were finishing our work, the preprint \cite{shen2025covariance} appeared. It uses a DMFT ansatz with random matrix techniques to study correlations of the covariance matrix of nonlinear networks.

This theoretical analysis requires modeling of the dynamics incorporating noisy contributions and the appropriate time scales. In the continuous-time recurrent neural network equation that we will use, there are two relevant time scales: the neuron time constant $\tau$, and the noise correlation time $\tau_\textrm{noise}$. The internal noise is often taken to be Gaussian and white, corresponding to $\tau_\textrm{noise}/\tau \to 0$. This is a useful mathematical simplification, but it is not always very realistic since it corresponds to an internal random variable that is changing infinitely fast. In this work we will instead take a complementary idealized limit: $\tau_\textrm{noise} \gg \tau$. In statistical mechanics terms, this corresponds to quenched disorder, as opposed to the annealed disorder that arises for $\tau_\textrm{noise} \to 0$. We will find that this provides important advantages for our goal of obtaining the large $N$ neural correlators. More realistic systems have $\tau \sim \tau_\textrm{noise}$, and this intermediate regime is harder to study theoretically. Nevertheless, it is interesting that the results that we obtain in the quenched limit $\tau_\textrm{noise} \gg \tau$ are not that different from those for $\tau_\textrm{noise} \ll \tau$. In this way, the quenched analysis of our work complements previous works on the white noise limit (which we will review below), and strongly suggests that the more realistic intermediate regime may after all not be that distinct from either of the mathematically simplified limits.

The work is organized as follows. In Sec. \ref{sec:setup} we introduce the model, the path integral description and the collective fields. In Sec. \ref{sec:largeN} we evaluate the large $N$ limit via a saddle point approximation. This section contains our main theoretical results: the generating function for the connected correlations of the network. This can be used to derive all correlation functions; we give explicitly the correlators for the covariance matrix. Using this, we derive an explicit and general expression for the dimension of participation. 

Sec. \ref{sec:appl} contains our main numerical results, exhibiting various concrete applications of our approach. First we consider power-law activation functions, and derive analytic results for the covariance matrix statistics. Then we introduce a class of ``Padé activation'' functions, which have various useful theoretical properties. We compute the analytic predictions and perform detailed comparisons with numerical simulations, finding excellent agreement. We discuss $1/N$ corrections and non-equilibrium effects. We compare our results with previous ones in the literature, including work on linear networks, as well as more recent results on nonlinear RNNs in the complementary white noise regime.

Finally, Sec. \ref{sec:concl} is devoted to the conclusions and future directions. While the main text focuses on the correlations of neural outputs, for completeness in Appendix \ref{app:input} we show how to derive the correlations for the inputs as well. Appendix \ref{app:numerics} is devoted to the numerical methods. Appendix \ref{app:deriv1} presents in more detail some steps of our analytical results.

\section{Model and path integral representation}\label{sec:setup}

A Recurrent Neural Network (RNN) for $N$ neurons with signal variable $\phi_i$, $i=1,\ldots, N$, and nonlinear activation function $f(\phi)$, is described by the equation \footnote{Equation~\eqref{eq:eom11} is a standard continuous-time firing-rate (leaky-integrator) network
driven by additive Gaussian noise, i.e. a Langevin equation. In the special linear case
$f(x)=x$ (or after linearization around a fixed point), it reduces to a multivariate
Ornstein--Uhlenbeck (linear Langevin) process.
See e.g. \cite{ginzburg1994theory, dahmen2020strong, doi:10.1073/pnas.1818972116, PhysRevLett.61.259} for closely related rate-network dynamics and correlation theories.}
\be\label{eq:eom11}
\tau \frac{d\phi_i(t)}{dt} = -\phi_i(t) + \sum_{j=1}^N W_{ij} f(\phi(t))_j+ \xi_i(t)\,,
\ee
where $\phi$ describes the input neural current, and $f(\phi)$, the activation function, gives the neural outputs. 
The number of neurons $i=1, \ldots, N$ is taken to be large.
We assume that the nonlinear function $f$ is regular, odd under $\phi \to -\phi$ and vanishes at the origin,
\be
f(\phi) = \phi + f_{(1)} \phi^3+ \ldots
\ee
This allows to expand the dynamics of $\phi$ around the fixed point $\phi=0$. The coefficient of the linear term can be set to one by a rescaling of $W$. The following results can be generalized to activation functions without parity and cases where $\phi$ acquires an expectation value. But we will restrict our analysis to this simpler case.

The random connectivity matrix usually fluctuates on time-scales which are much larger than typical single-neuron time variations. So we will take it as a quenched disorder variable described by a Gaussian noise with vanishing expectation value and
\be
\langle W_{ij} W_{kl} \rangle = \frac{\lambda^2}{N} \delta_{ik} \delta_{jl}\,.
\ee
The factor of $N$ in this variance is chosen to have a well-defined large $N$ limit\footnote{Let us note that this distribution is unrealistic from a biological point of view. Real weight distributions are log-normal instead of Gaussian and they also display significant correlations \cite{song2005highly}}.
On the other hand, the random internal noise $\xi(t)$ can be modeled by its own Ornstein-Uhlenbeck with time-scale $\tau_\textrm{noise}$,
\be
\frac{d \xi_i(t)}{dt} = - \frac{\xi_i(t)}{\tau_\textrm{noise}}+ \sigma \eta_i(t)
\ee
with $\eta_i(t)$ a random Gaussian white noise variable, $\langle \eta(t) \eta(t') \rangle=\delta(t-t')$. At late times then,
\be\label{eq:OUxi}
\langle \xi_i(t) \xi_j(0) \rangle \approx  \delta_{ij} \,\frac{\sigma^2 \tau_\textrm{noise}}{2} e^{-|t|/\tau_\textrm{noise}}\,.
\ee

Experimentally, the timescale associated with input synchrony, which motivates our $\tau_\textrm{noise}$, can range from a few tens to a few hundred milliseconds, approximately $20$--$200$ ms, depending on whether the cortical network is in a spontaneous state or in a stimulus-driven state \cite{pattadkal2026synchrony}. By contrast, intrinsic cortical neuronal timescales, related with the dynamical timescale $\tau$, are typically of order $10$--$20$ ms \cite{mccormick1985comparative}. The annealed (white-noise) limit corresponds to $\tau_\textrm{noise}\to 0$, so that the time dependence in (\ref{eq:OUxi}) becomes proportional to $\delta(t)$, keeping $\sigma^2\tau_\textrm{noise}$ fixed. This limit has been analyzed before in our context by e.g. \cite{doi:10.1073/pnas.1818972116,hu2022spectrum,shen2025covariance}. In this work we will instead focus on the limit $\tau \ll \tau_\textrm{noise}$, and dynamics with times $t$ such that $\tau \ll t \ll \tau_\textrm{noise}$. This means that the equilibrium configurations and their corresponding correlation functions can be obtained while keeping the random variable $\xi$ approximately constant (quenched disorder). This corresponds to taking
\be
\langle \xi_i \xi_j\rangle= \delta_{ij} D\,,
\ee
with $D \sim \sigma^2 \tau_\textrm{noise}$ fixed.

\subsection{Equilibrium partition function representation}

We will focus for simplicity on equilibrium correlation functions. In the regime $\tau \ll  \tau_\textrm{noise}$, the equilibrium configuration is described by
\be\label{eq:eomeq}
0 = -\phi_i + \sum_j\,W_{ij} f(\phi_j)+ \xi_i\,,
\ee
with $W$ and $\xi$ random quenched disorder variables as discussed before.  This will provide us with important mathematical simplifications that will lead to explicit results in the large N limit. Below we will also compare with the recent results of \cite{shen2025covariance} in the opposite limit $\tau_\textrm{noise} \ll \tau$.  It would be very interesting to extend the analysis below to include the time-dependent dynamics of the network, something that we plan to address in future work. The numerical simulation of Eq. (\ref{eq:eom11}) is described in App. \ref{app:numerics}.

Our goal is to compute the equilibrium correlation functions of the network outputs,
\be\label{eq:GammaNr}
\Gamma^{N_r}_{i_1 \ldots, j_1,\ldots,s_1\ldots}= \langle \langle f(\phi_{i_1}) f(\phi_{i_2}) \ldots \rangle_\xi  \langle f(\phi_{j_1}) f(\phi_{j_2}) \ldots \rangle_\xi \ldots \langle f(\phi_{s_1}) f(\phi_{s_2}) \ldots \rangle_\xi  \rangle_W\,,
\ee
Here $\langle \ldots \rangle_\xi$ denotes the average over $\xi$, $\langle \ldots \rangle_W$ is the average over $W$, and $N_r$ is the number of $\xi$-averages inside the $\langle \ldots \rangle_W$. The calculation of the correlations of the network inputs $\phi_i$ can be done along similar lines; this is presented in the Appendix \ref{app:input}.

It is important to notice that experimental evaluations of the correlations do not involve an average over the coupling variables, but rather empirical averages over some sets of neurons. Here we are assuming that in the limit of large $N$  the macroscopic state does not depend on the precise value of each of their random parameters but rather only on their statistics; see also \cite{doi:10.1073/pnas.1818972116} for more discussions on this point.  In other words we assume the system is {\it self-averaging}. This assumption  allows us to use the average over $W$ to predict the empirical averages.

For the purpose of obtaining the large $N$ correlators, we will find it convenient to proceed in terms of a partition function representation. 
Consider first a single $\xi$-average. By the assumed parity property $f(-\phi)=-f(\phi)$, correlation functions with an odd number of insertions vanish. Correlation functions with an even number of $f$-insertions can be obtained from the partition function with a source $J_{ij}$ for $f_i  f_j$.
Introducing a Lagrange multiplier field $\t \phi_i$ that imposes the equilibrium stochastic equation (\ref{eq:eomeq}), the integral representation for the generating function including the $\xi$-noise average is
\begin{align}
Z_{\xi}(J)
&= \int \mathcal{D}\xi\,\mathcal{D}\phi\,\mathcal{D}\tilde{\phi}\;F(W, \xi)\;
e^{-\frac{1}{2D}\sum_i \xi_i^2} 
e^{
i\sum_i \tilde{\phi}_i\left(\phi_i-\sum_j W_{ij}\,f(\phi_j)-\xi_i
\right)} 
e^{
\sum_{ij}
J_{ij}\,f(\phi_i)\,f(\phi_j)}\,.
\label{eq:Zxi1}
\end{align}
We denote ${\mathcal D} \phi = \prod_i d\phi_i$, and similarly for other integrated variables.
Here $F(W, \xi)$ arises from the Jacobian of the delta function produced by the Lagrange multiplier.\footnote{In other words, for the integral over all possible $\phi$, $F$ satisfies $\delta(\phi-\phi_{\textrm{sol}})=F(W,\xi) \delta(\phi_i-\sum_j W_{ij}\,f(\phi_j)-\xi_i)$, where $\phi_{\textrm{sol}}$ is the solution to \ref{eq:eomeq}.} In the linear model, a short calculation gives $F(W,\xi)$ as a determinant of the $\phi$ and $\t \phi$ Green's function, but more generally it is a complicated function that depends on the activation function and cannot be computed in closed form. We will see shortly that its effects are subleading for a large network.

Now let's turn to the generating function of $W$-averages of the form (\ref{eq:GammaNr}). Each $\xi$-average is represented by a generating function (\ref{eq:Zxi1}), so we need to introduce $N_r$ replicas; the replica indices will be denoted by $a,b,\ldots= 1, \ldots, N_r$. Then the fields in (\ref{eq:Zxi1}) acquire two indices: $\xi_i^a,\,\phi_i^a,\, \t \phi_i^a, \,f_i^a\equiv f(\phi_i^a)$, while $W_{ij}$ has no replica indices because we are considering a single $W$-average. Furthermore, note that by parity only correlation functions of $f_i^a$ with an even number of $i$-indices are nonvanishing. In order to generate these, it is sufficient to use a bi-source $J_i^{ab}$ that is diagonal in physical space (neuron indices $i$) and nondiagonal in replica space. For instance,
\be
\langle \langle f_1 f_2 \rangle_\xi \langle f_1 f_2 \rangle_\xi \rangle_W = \langle f^1_1 f^1_2  f^2_1 f^2_2  \rangle =\langle f^1_1f^2_1 f^1_2   f^2_2  \rangle= \frac{1}{Z_{N_r=2}(0)} \frac{\partial}{\partial J_{1}^{12}} \frac{\partial}{\partial J_{2}^{12}} Z_{N_r=2}(J) \Big |_{J=0}\,.
\ee
Therefore, the generating function for the statistics of $N_r$ output correlations is
\be\label{eq:Zxi}
Z_{N_r}(J)= \int \mathcal{D} W\, \mathcal{D}\xi\,\mathcal{D}\phi\,\mathcal{D}\tilde{\phi}\; F(W,\xi)^{N_r}\;e^{-S}\,,
\ee
with action
\bea
S
&=&\frac{N}{2\lambda^2}\sum_{i,j=1}^{N} W_{ij}^2
+\sum_{i=1}^{N}
\sum_{a=1}^{N_r}\left[
\frac{1}{2D}\big(\xi_i^a\big)^2
-i\tilde\phi_i^a\Big(\phi_i^a-\sum_{j=1}^{N} W_{ij}f_j^a-\xi_i^a\Big)
\right] \nonumber\\
&&\qquad-\sum_{i=1}^{N}\sum_{a,b=1}^{N_r} J_i^{ab}\,  f_i^a\,f_i^b\,.
\label{eq:Sstatic}
\eea

As before, the notation $\mathcal{D} X$ stands for the products of all the differentials in the $X$ vector or matrix; so, for example $\mathcal{D} W =\prod_{ij} dW_{ij}$, $\mathcal D \phi= \prod_{i,a} d\phi_i^a$ and so on.
The external source $J_i^{ab}$ is chosen to be an upper-triangular matrix in replica space.\footnote{We could also choose it to be symmetric in the indices $(a,b)$, at the cost of introducing factors of $2$ between the source derivatives of $Z(J)$ and the correlation functions.}
As long as $N_r \ll N$, the factor of $F(W,\xi)^{N_r}$ gives a small correction to the Gaussian dynamics of $W$ and $\xi$, and so we will neglect it in what follows.\footnote{More explicitly, the Jacobian factor is proportional to $| det(1- W f' ) | = | e^{\sum_i \log(1- W_{ii} f_i')}|$. It only includes the $N$ diagonal terms. This is explicitly subleading compared to the order $N^2$ terms in the action for $W$, $\sum_{ij} W_{ij}^2$, which moreover in the large $N$ normalization has an explicit $N$ in front.
}

Performing the Gaussian $\xi$ and $W$-integrals gives
\be
Z_{N_r}(J)=\int \mathcal{D} \phi\,\mathcal{D} \tilde \phi\;
\exp\!\left[\sum_{i=1}^{N}\left(
-\frac12\sum_{a,b=1}^{N_r} M_{ab}\,\tilde\phi_i^a\tilde\phi_i^b
+i\sum_{a=1}^{N_r}\phi_i^a\tilde\phi_i^a
+\sum_{a,b=1}^{N_r}J_i^{ab}\, f_j^a\,f_j^b
\right)\right]
\ee

where we have defined
\be
M_{ab}
= \delta_{ab} \, D+\frac{\lambda^2}{N}\sum_{j=1}^{N} f_j^a\,f_j^b.
\ee
We see that the effect of the random variable $\xi_i$ is to produce a quadratic term for $\t \phi_i$ proportional to its variance $D$, while the Gaussian noise $W$ introduces interactions with coupling $\lambda^2/N$. When the activation function is well approximated by a linear regime, this gives a $\phi^4$-type interaction. In this work we will focus on more general nonlinear activation functions. This normalization with $N$ is the natural one in order to produce finite correlation functions when $N \to \infty$; see e.g. \cite{Moshe:2003xn} for a review on the $1/N$ expansion.
Next, the Gaussian $\t \phi$ integral gives
\begin{align}
Z_{N_r}(J)
&=\int  \mathcal{D} \phi\;
\big[\det(2\pi M^{-1})\big]^{N/2}
\exp\!\left[
\sum_{i=1}^{N}\sum_{a,b=1}^{N_r}\left(
-\frac12 (M^{-1})_{ab}\,\phi_i^a\phi_i^b
+J_i^{ab}\,f_i^a f_i^b
\right)
\right].
\label{eq:Znr0}
\end{align}

\subsection{Collective fields}\label{subsec:coll}

In order to perform a large $N$ expansion we need to introduce 
 collective fields so that the path integral becomes dominated by a ``classical'' saddle point. The model (\ref{eq:Znr0}) is a particular case of a vector model, and the appropriate collective coordinate is \cite{Moshe:2003xn} 
\be\label{eq:rhocoll}
\rho_{ab}= \frac{1}{N} \sum_j f_j^a f_j^b\,.
\ee
Note that this is a symmetric matrix in the replica indices.
 Introducing an auxiliary field $\eta_{ab}$ that enforces this relation as a constraint ($\eta_{ab}$ is also symmetric in $(a,b)$), we arrive at the following representation for the path integral:
\begin{align}
Z_{N_r}(J)
&=\int {\mathcal D}\phi\,  {\mathcal D}\eta\,  {\mathcal D}\rho\;
\exp\!\left[
-\frac{i}{2}\sum_{a,b=1}^{N_r}\eta_{ab}\Big(\sum_{j=1}^{N} f_j^a f_j^b - N\rho_{ab}\Big)
\right]\,
\big[\det(2\pi M^{-1})\big]^{N/2}
\nonumber\\
&\qquad\times
\exp\!\left[
\sum_{i=1}^{N}\sum_{a,b=1}^{N_r}\left(
-\frac12 (M^{-1})_{ab}\,\phi_i^a\phi_i^b
+J_i^{ab}\,f_i^a f_i^b
\right)
\right].
\label{eq:Znr1}
\end{align}

The key advantage of this representation is that now
\be\label{eq:Mnew}
M_{ab}=\delta_{ab} D+ \lambda^2 \rho_{ab}\,,
\ee
is independent of $\phi$. The equivalence with (\ref{eq:Znr0}) follows from the fact that the integral over  $\eta$ creates a delta function enforcing (\ref{eq:rhocoll}). These fields will play the role of collective coordinates describing the correlations of the RNN.

The integrals appearing here cannot be obtained in general due to the nonlinear activation function. However, our main result will be that, when $N \gg 1$, a semi-classical saddle point configuration controls the partition function. This will allow us to evaluate correlation functions analytically in a $1/N$ expansion, thus providing a new nonperturbative framework for understanding the dynamics of nonlinear RNNs. For related work on path integral representations with collective fields see also \cite{PhysRevX.8.041029, doi:10.1073/pnas.1818972116}.

\section{Large $N$ solution}\label{sec:largeN}

The next step is to extract the quadratic dependence on $\phi$ from the activation function and then perform the $\phi$ integral. For this, we define a new function
\be
g_{ab}(\phi_j)\equiv f_j^a f_j^b- \phi_j^a \phi_j^b\,
\ee
and so
\begin{align}
Z_{N_r}(J)
&=\int  {\mathcal D}\phi\,  {\mathcal D}\eta\,  {\mathcal D}\rho\;
\exp\!\left[
-\frac{i}{2}\sum_{a,b=1}^{N_r}\eta_{ab}\left(\sum_{j=1}^{N} g_{ab}(\phi_j)-N\rho_{ab}\right)
\right]\,
\big[\det(2\pi M^{-1})\big]^{N/2}
\nonumber\\
&\qquad\times
\exp\!\left[
\sum_{j=1}^{N}\sum_{a,b=1}^{N_r}\left(
-\frac12 (G^{-1})_{ab}\,\phi_j^a\phi_j^b
+J_{j}^{ab}\,\big(g_{ab}(\phi_j)+\phi_j^a\phi_j^b\big)
\right)
\right]
\label{eq:Znr2}
\end{align}
where
\be\label{eq:Ginvdef}
G^{-1}_{ab}=  (M^{-1})_{ab}+ i \eta_{ab}\,.
\ee

In the large $N$ limit with the number of sources fixed, the contributions from the source terms are suppressed by $1/N$ compared to the other terms in (\ref{eq:Znr2}). Therefore, in the limit of a large number of neurons, the highly complicated correlations of the neural system can be mapped into a dual classical configuration for the ``master fields'' $\rho$ and $\eta$, which extremize the integrand of (\ref{eq:Znr2}). The existence of such dual descriptions is one of the key advantages of large $N$ systems, and has been very influential in the development of quantum field theory and gravity \cite{Aharony:1999ti}, and more recently also in deep learning \cite{Roberts:2021fes}. Here we also find that it plays a central role in the understanding of nonlinear RNNs.

\subsection{$1/N$  and source expansion}\label{subsec:Nlarge}

At large $N$, the sources are then small perturbations of a network that is homogeneous, namely each neuron site $j$ has the same dynamics. Let us denote a single neuron field $\phi_j^a$ by the variable $x^a$, and introduce the  normalized Gaussian average
\be\label{eq:Gexp}
\langle F(x) \rangle_G \equiv \int \frac{d^{N_r}x}{[\det(2\pi G)]^{1/2}}\,F(x) e^{-\frac{1}{2} \sum_{a,b=1}^{N_r}(G^{-1})_{ab}x^a x^b} \,.
\ee
The partition function then becomes
\begin{align}
Z_{N_r}(J)
&=\int  {\mathcal D}\eta\,  {\mathcal D}\rho\;
\exp\!\left[\frac{i}{2}N\,\tr(\eta\rho)\right]\,
\big[\det(1+iM\eta)\big]^{-N/2}
\nonumber\\
&\qquad\times
\prod_{j=1}^{N}
\left\langle
\exp\!\left[
-\frac{i}{2}\tr\!\big(\eta\, g(x)\big)
+\tr\!\big(J_j\,(g(x)+xx)\big)
\right]
\right\rangle_G .
\label{eq:Znrnoexpand}
\end{align}
In order to shorten some of the following formulas, we will use a notation where $xx$ is a matrix with elements $(xx)^{ab}= x^a x^b$, and $\tr (J_j x x) =\sum_{ab} J_j^{ab} x^a x^b$.

Our goal is to compute the partition function including second order contributions from the sources, and the dominant terms at large $N$. Let us write this as
\be\label{eq:Znrfinal}
Z_{N_r}(J) = \int  {\mathcal D}\eta\,  {\mathcal D}\rho\, e^{-N S_{eff}}\,.
\ee
The calculation of the  ``effective action'' (in a statistical mechanics sense) is demonstrated in detail in Appendix \ref{app:deriv1}, and here we summarize the main steps. The first step is to write a cumulant expansion for the factor of (\ref{eq:Znrnoexpand}) containing the sources. This can be done to all orders in powers of $\eta g$. The second step involves taking the large $N$ limit, and here comes a crucial simplification: it is possible to  check self-consistently that the diagonal components of $\rho_{ab}$ and $G_{ab}$ are order $N^0$, while the off-diagonal components of $\rho_{ab}$ and $G_{ab}$, as well as all the variables $\eta_{ab}$ are order $1/N$. With this $N$-scaling, the leading terms at large $N$ are
\begin{align}\label{eq:Seff2}
S_{eff}&\approx \frac{1}{2} \log \det(1+i M \eta)-\frac{i}{2}\tr( \eta \rho)+ \frac{i}{2} \tr\left(\eta \langle g \rangle_G\right)+\frac{1}{2} \sum_{a<b}\eta_{ab}^2\langle g_{ab}^2 \rangle_G  \\
&-\frac{1}{N}\sum_j \left[ \sum_a J_j^{aa} \left(G_{aa}+ \langle g_{aa}\rangle_G\right) + \sum_{a<b} J_j^{ab}\left( G_{ab}+ \langle g_{ab}\rangle_G - i \eta_{ab} \langle g_{ab}(g_{ab}+x^a x^b )\rangle_G\right)\right]\notag\\
&-\frac{1}{2N}\sum_j \left[ \sum_a  (J_j^{aa})^2 \left(\langle(g_{aa}+x^a x^a)^2 \rangle_G-\langle g_{aa}+x^a x^a \rangle_G^2 \right) +\sum_{a<b} (J_j^{ab})^2 \langle (g_{ab}+ x^a x^b)^2 \rangle_G \right]\notag\,.
\end{align}

We have introduced the off-diagonal contribution $\sum_{a<b} \eta_{ab}^2\langle g_{ab}^2 \rangle_G $ because it turns out to contribute at order $1/N^2$; this will have the same $N$-scaling as  $\eta_{ab} \langle g_{ab} \rangle_G$, which is the leading contribution for $a \neq b$. In contrast, the leading contribution from $\eta_{ab} \langle g_{ab} \rangle_G$ for $a=b$, occurs at order $1/N$, while the corresponding quadratic term $ \eta_{aa}^2 \langle g_{aa}^2 \rangle_G \sim 1/N^2$ and hence can be neglected. The same $N$-scaling arguments have been used for the remaining terms we kept in \eqref{eq:Seff2}.

\subsection{Effects of nonlinearities}\label{subsec:nonlinear}

Let us now incorporate more explicitly the effects from the nonlinearity of the network, encoded in the insertions of $g_{ab}$ in (\ref{eq:Seff2}).

In order to evaluate the $G$-expectation values in the effective action, we will approximate $G^{-1}$ in (\ref{eq:Ginvdef}), taking into account that $G_{aa}\sim N^0$ while $G_{ab} \sim N^{-1}$ if $b \neq a$.\footnote{At $J=0$, different replicas are identical but independent at leading order; the only source of replica
coupling is the disorder average, which enters with the $1/N$ scaling of the weight covariance.
Therefore the replica off-diagonal components are expected to vanish at leading order and to appear
first as $O(1/N)$ corrections, a scaling that we will verify self-consistently at the saddle point.
} To leading order in the off-diagonal entries, this yields

\be \label{Gapprox}
(G^{-1})_{aa} \approx \frac{1}{G_{aa}}\;,\;(G^{-1})_{a\neq b} \approx - \frac{G_{ab}}{G_{aa} G_{bb}}\,.
\ee

As before, sums over replica indices are always written explicitly. With this approximation,
\be
\langle g_{aa}\rangle_G \approx \int \frac{dx}{\sqrt{2\pi G_{aa}}}(f^a f^a- x^a x^a) e^{-\frac{1}{2} G_{aa}^{-1}(x^a)^2} = G_{aa} \left(V(G_{aa})-1 \right)\,,
\ee
where we have defined
\be\label{eq:Vdef}
V(G) \equiv \frac{1}{G}\, \int \frac{dx}{\sqrt{2\pi G}} f(x)^2\, e^{-\frac{x^2}{2G}}\,.
\ee

For the off-diagonal component the leading nonvanishing contribution comes from expanding
the Gaussian weight to first order in $G_{ab}$. In particular, using \ref{Gapprox} one has a cross-term
$\exp\!\big[(G_{ab}/(G_{aa}G_{bb}))x^a x^b\big]\simeq 1+(G_{ab}/(G_{aa}G_{bb}))x^a x^b$.
For odd activation functions (more generally, whenever $\langle f(x)\rangle_G=0$), the zeroth-order term
vanishes and we obtain
\begin{align}
\langle g_{a\neq b}\rangle_G
&\approx \int \frac{dx^a\,dx^b}{\sqrt{(2\pi G_{aa})(2\pi G_{bb})}}\,
\Big(f^af^b-x^a x^b\Big)\,
\left(\frac{G_{ab}}{G_{aa}G_{bb}}\right) x^a x^b\,
e^{-\frac{(x^a)^2}{2G_{aa}}-\frac{(x^b)^2}{2G_{bb}}}\\
&= \frac{G_{ab}}{G_{aa}G_{bb}} \left( \langle xf(x) \rangle_{G_{aa}} \langle xf(x) \rangle_{G_{bb}} \right) -G_{ab}
\end{align}
We introduce the function
\be\label{eq:Udef}
U(G)  \equiv \frac{1}{G} \int \frac{dx}{\sqrt{2\pi G}}\,x f(x)\,e^{- \frac{x^2}{2G}}\,,
\ee
and then the off-diagonal nonlinearity contribution becomes
\be
\langle g_{a \neq b}\rangle_G \approx G_{ab} \left[U(G_{aa}) U(G_{bb})-1 \right]\,.
\ee

Applying the same first-order approximation for the Gaussian weight, we obtain
\be
\langle g_{a \neq b}^2 \rangle_G
= \langle g_{a\neq b}^2 \rangle_0
+ \frac{G_{ab}}{G_{aa}G_{bb}} \langle g_{a\neq b}^2 x^a x^b \rangle_0
+ \mathcal{O}(G_{ab}^2)
\approx \langle g_{a\neq b}^2 \rangle_0,
\ee
where $\langle\cdot\rangle_0$ denotes the decoupled Gaussian average. Expanding the integrand and using the factorization of the decoupled measure, we find
\be
\langle g_{a \neq b}^2 \rangle_G
\approx
\langle f(x)^2 \rangle_{G_{aa}} \langle f(x)^2 \rangle_{G_{bb}}
- 2\langle x f(x) \rangle_{G_{aa}} \langle x f(x) \rangle_{G_{bb}}
+\langle (x^a)^2\rangle_{G_{aa}} \langle (x^b)^2\rangle_{G_{bb}}\,.
\ee
Finally, applying the definitions in Eqs.~\ref{eq:Vdef} and \ref{eq:Udef}, we obtain
\be
\langle g_{a\neq b}^2\rangle_G \approx G_{aa} G_{bb}\left[V(G_{aa}) V(G_{bb})-2 U(G_{aa}) U(G_{bb})+1 \right]\,.
\ee
A similar calculation gives
\be
\langle g_{a\neq b} x^a x^b \rangle_G \approx G_{aa} G_{bb}\left[U(G_{aa}) U(G_{bb})-1 \right]\,.
\ee

Substituting the explicit Gaussian averages computed above into Eq.~\eqref{eq:Seff2}, and using the decomposition for symmetric replica matrices $\tr(\eta A)=\sum_a \eta_{aa}A_{aa}+2\sum_{a<b}\eta_{ab}A_{ab}$, we rewrite each term in $S_{\rm eff}$. The combination appearing in the $\mathcal{O}(J)$ term becomes 
\begin{align} \langle g_{ab}(g_{ab}+x^a x^b)\rangle_G =\langle g_{ab}^2\rangle_G+\langle g_{ab}\,x^a x^b\rangle_G \approx G_{aa}G_{bb}\Big(V(G_{aa})V(G_{bb})-U(G_{aa})U(G_{bb})\Big)
\end{align}

The last function we need to introduce comes from
\be
\langle(g_{aa}+x^a x^a)^2 \rangle_G = G_{aa}^2 (W(G_{aa})+3)\,,
\ee
namely
\be
W(G_{aa}) =\frac{1}{G_{aa}^2} \left( \langle g_{aa}^2 \rangle_G+2 \langle g_{aa}x^a x^a\rangle_G \right)\,.
\ee


In this way, we arrive at our final form for the effective action,
\begin{align}\label{eq:Seff3}
S_{eff}&\approx \frac{1}{2} \log \det(1+i M \eta)-\frac{i}{2}\sum_a \eta_{aa} \left[\rho_{aa}- G_{aa} \left(V(G_{aa})-1 \right)\right]\nonumber\\
&-i\sum_{a<b} \eta_{ab} \left(\rho_{ab}- G_{ab} \left[U(G_{aa}) U(G_{bb})-1 \right]\right)\\
&+\frac{1}{2} \sum_{a<b}\eta_{ab}^2 G_{aa} G_{bb}\left[V(G_{aa}) V(G_{bb})-2 U(G_{aa}) U(G_{bb})+1 \right] -\frac{1}{N}\sum_a (\sum_j J_j^{aa})  G_{aa} V(G_{aa}) \nonumber \\
&-\frac{1}{N} \sum_{a<b}(\sum_j J_j^{ab})\left( G_{ab} U(G_{aa})U(G_{bb})- i \eta_{ab} G_{aa} G_{bb}\left[V(G_{aa})V(G_{bb})-U(G_{aa})U(G_{bb}) \right]\right)\nonumber\\
&-\frac{1}{2N}\sum_j \left [ \sum_a  (J_j^{aa})^2\, G_{aa}^2 (W(G_{aa})-V(G_{aa})^2+3)+\sum_{a<b} (J_j^{ab})^2\,G_{aa} G_{bb} V(G_{aa}) V(G_{bb})\right ]\nonumber\,.
\end{align}
The functions $V$, $U$ and $W$ are Gaussian integrals containing the nonlinear activation function, while $M$ and $G$ are given in terms of $\rho$ and $\eta$ in Eqs. (\ref{eq:Mnew}) and (\ref{eq:Ginvdef}).

\subsection{Saddle point description of neural correlation functions}\label{subsec:saddle}

At large $N$, (\ref{eq:Znrfinal}) is dominated by the saddle point solution,
\be
Z_{N_r}(J)\approx e^{-N S_{eff,saddle}(J)}\,,
\ee
where the saddle point for the collective coordinates is determined by
\be
\frac{\partial S_{eff}}{\partial \rho_{ab}}=\frac{\partial S_{eff}}{\partial \eta_{ab}}=0\;.
\ee
We will solve these equations in a $1/N$ expansion:
\be
\rho_{ab}= \rho_{ab}^{(0)} + \frac{1}{N} \rho_{ab}^{(1)}+\ldots\;,\;\eta_{ab}= \eta_{ab}^{(0)} + \frac{1}{N} \eta_{ab}^{(1)}+\ldots
\ee

At order $N^0$, the result is
\be
\rho_{ab}^{(0)} = \delta_{ab} \rho_0\;,\; \eta_{ab}^{(0)}=0\,,
\ee
where $\rho_0$ is the solution to
\be\label{eq:rhoself}
\left(D_0+\lambda ^2 \rho_0\right) V\left(D_0+\lambda ^2  \rho_0\right)-\rho_0=0\,.
\ee
The fact that $\eta=0$ at leading order provides a key simplification for the collective field description of nonlinear networks: it allows to perform a perturbative expansion of the interaction term $\langle e^{-(i/2) \eta g}\rangle_G$ in (\ref{eq:Znrnoexpand}), as we did in (\ref{eq:Seff2}).\footnote{A similar simplification was observed for linear networks in \cite{doi:10.1073/pnas.1818972116}.} At this order, all the effects from the nonlinearities are then encoded in the self-consistent equation (\ref{eq:rhoself}) for the collective field $\rho$. From (\ref{eq:Znr2}) and (\ref{eq:Ginvdef}), the two-point function for the input variables at order $N^0$ becomes 
\be
\langle \phi_i^a \phi_j^b \rangle_{W, \xi} =\delta_{ij} \delta_{ab} G_0
\ee
with 
\be
G_0 = D+ \lambda^2 \rho_0\,.
\ee
We will find it convenient to express our results in terms of this leading 2-point function $G_0$. Replacing $\rho_0$ in terms of $G_0$, we find that the self-consistent equation (\ref{eq:rhoself}) becomes
\be\label{eq:Gself}
G_0=D+\lambda ^2\,G_0 V(G_0)\,.
\ee
Recall that the function $V(x)$ is determined in terms of the activation function via (\ref{eq:Vdef}). We will give explicit examples below. In the cases of interest, this self-consistent equation shows that $G_0(\lambda, D)$ is a monotonic function of the coupling $\lambda$. It is then simpler to just use (\ref{eq:Gself}) to write $\lambda^2$ as a function of $G_0$, and this is what we use for the analytic results in the next section.

It is also useful to rewrite this self-consistent equation in terms of
\be\label{eq:Cphidef}
\langle C^\phi \rangle_W = \langle \phi_i \phi_i \rangle_{\xi, \,W}=G_0\;,\;\langle C^f \rangle_W = \langle f_i(\phi) f_i(\phi) \rangle_{\xi, \,W}=G_0 V(G_0)\,.
\ee
Then (\ref{eq:Gself}) becomes
\be\label{eq:selfCphi}
\langle C^\phi \rangle_W= D + \lambda^2 \,\langle C^f \rangle_W\,,
\ee
which we recognize as a mean field equation. Such expressions have appeared before, e.g. in \cite{PhysRevLett.61.259, PhysRevX.8.041029}.  It has the same structure as the self-consistent equation (2.2) found very recently in \cite{shen2025covariance}; we will compare with their results in Sec. \ref{subsec:comparison}.

For the linear network, $f(x)=x$, $V(G)=1$, and (\ref{eq:Gself}) can be solved to give
\be
G_0= \frac{D}{1-\lambda^2}\,.
\ee
This result has been previously reported in \cite{doi:10.1073/pnas.1818972116}. Therefore, the 2-point function diverges as the coupling $\lambda \to 1$. This limit corresponds to the instability of the linear network, where the linear restoring term in (\ref{eq:eom11}) is overcome by the effects from fluctuations of the connectivity matrix. Understanding the network dynamics beyond this point requires including effects beyond the linear approximation. In our present formalism, we are including nonlinear effects exactly in the $1/N$ approximation; this divergence is resolved for activation functions that are bounded by $f(x)< x^p$ for $p<1$ at large $|x|$. To see this, Eq. (\ref{eq:Gself}) can be solved at large $G_0$ approximating the integral in $V(G_0)$  by the regime where $f(x) \sim x^p$; the self-consistent solution then shows that the growth of $G_0$ is bounded by $G_0 \lesssim  (\textrm{const})\,\lambda^{2/(1-p)}$. See Sec \ref{subsec:power-law} for more details.

At the next order in the $1/N$ expansion, the collective fields $\rho_{ab}^{(1)}$ and $\eta_{ab}^{(1)}$ are proportional to linear and quadratic combinations of the sources. In other words, at order $1/N$ the collective fields encode the backreaction of the insertions of sources and the interplay with the network nonlinearity. Let us exhibit these effects explicitly at linear order in the sources:
\begin{align}
\rho_{aa}^{(1)}&= \frac{2 G_0^2 \left(G_0 V'(G_0)+V(G_0)\right) \left(\lambda ^2 \left(G_0 V'(G_0)+V(G_0)\right)-\lambda ^2+1\right)}{\left(\lambda ^2 \left(G_0 V'(G_0)+V(G_0)\right)-1\right)^2}\,\sum_j J^{aa}_j + \mathcal O(J^2)\nonumber\\
\rho_{a b}^{(1)}&=\frac{G_0^2 V(G_0)^2}{\left(\lambda ^2 U(G_0)^2-1\right)^2}\sum_j J^{ab}_j + \mathcal O(J^2)\;,\;a \neq b\,,
\end{align}
and
\begin{align}
\eta_{aa}^{(1)}&= -2i \left(\frac{1}{\lambda ^2 \left(G_0 V'(G_0)+V(G_0)\right)-1}+1\right)\,\sum_j J^{aa}_j+ \mathcal O(J^2) \nonumber\\
\eta_{a b}^{(1)}&= i \frac{\lambda^2 U(G_0)^2 }{1-\lambda^2 U(G_0)^2}\sum_j J^{ab}_j + \mathcal O(J^2)\;,\;a \neq b\,.
\end{align}
The quadratic terms are straightforward to compute, we won't use their explicit expression.

Finally, computing the $1/N$ saddle point values $\rho_{ab}^{(1)}$ and $\eta_{ab}^{(1)}$ and replacing into the effective action, we obtain the generating function of connected correlation functions,
\begin{align}\label{eq:final1}
\log Z_{N_r}(J) &\approx  G_0 V(G_0) \sum_j \sum_a J_j^{aa}+ \frac{1}{2} G_0^2 V(G_0)^2 \sum_j \sum_{a<b} (J_j^{ab})^2 \nonumber\\
&+\frac{1}{2} G_0^2 (W(G_0)-V(G_0)^2+3) \sum_j \sum_a (J_j^{aa})^2\\
&+\frac{\lambda^2}{N}\frac{G_0^2 U(G_0)^2 V(G_0)^2 \left(2-\lambda^2 U(G_0)^2\right)}{2 \left(1-\lambda ^2 U(G_0)^2\right)^2} \sum_{a<b} (\sum_j J_j^{ab})^2 \nonumber\\
&+\frac{\lambda ^2}{N} \frac{G_0^2  \left(2-\lambda ^2\right) \left(G_0 V'(G_0)+V(G_0)\right)^2}{\left(1-\lambda^2 G_0 V'(G_0)+\lambda^2 V(G_0)\right)^2} \sum_a (\sum_j J_j^{aa})^2 \nonumber\,.
\end{align}
The first two lines in this expression have been simplified by using the self-consistent equation (\ref{eq:Gself}). The third and fourth lines can also be written in different ways by applying the self-consistent equation. For our results below we will use the form where $\lambda^2$ is solved for in terms of $G_0$ using (\ref{eq:Gself}). In particular, this will be useful for showing that $1-\lambda^2 U(G_0)^2$ is strictly positive, and can only vanish for a linear activation function.

Let us present these results in terms of neural correlation functions. We define the covariance matrix
\be
C_{ij}^f = \langle f_i(\phi) f_j(\phi) \rangle_\xi\,.
\ee
Then the output 2-point function at large $N$ becomes
\be\label{eq:C2point}
\langle C_{ij}^f \rangle_W =\delta_{ij} \frac{1}{Z_{N_r}} \frac{\partial Z_{N_r}}{ \partial J_i^{aa}} \Big|_{J=0} \approx \delta_{ij}\,G_0 V(G_0) \,.
\ee
From (\ref{eq:Vdef}), this is the Gaussian expectation value of $f(x)^2$, as expected. Let us now consider the 2-point functions of the covariance matrix $C_{ij}^f$. The first case is
\be
\langle C_{ii}^f C_{jj}^f \rangle_W= \frac{1}{Z_{N_r}} \frac{\partial^2  Z_{N_r}}{\partial J_i^{aa} \partial J_j^{bb}} \Big|_{J=0} \approx G_0^2 V(G_0)^2
\ee
where $i \neq j$ are some neuron indices, and $a \neq b$ are some replica indices. Therefore, these fluctuations approximately factorize at large $N$, 
\be\label{eq:factorize}
\langle C_{ii}^f C_{jj}^f \rangle_W \approx \langle C_{ii}^f\rangle_W \,\langle C_{jj}^f\rangle_W\,.
\ee 
We also find factorization if $i=j$:
\be\label{eq:factorize2}
\langle C_{ii}^f C_{ii}^f \rangle_W= \frac{1}{Z_{N_r}} \frac{\partial^2  Z_{N_r}}{\partial J_i^{ab} \partial J_i^{ab}} \Big|_{J=0} \approx G_0^2 V(G_0)^2 \approx \langle C_{ii}^f\rangle_W \,\langle C_{ii}^f\rangle_W
\ee

where again $a \neq b$ are two different replica indices. The same result is obtained by differentiating with respect to $J_i^{aa}$ and $J_i^{bb}$. 

On the other hand, the fluctuations of off-diagonal elements of the covariance matrix are
\be\label{eq:off}
\langle C_{ij}^f C_{ij}^f \rangle_W= \frac{1}{Z_{N_r}} \frac{\partial^2  Z_{N_r}}{\partial J_i^{ab} \partial J_j^{ab}} \Big|_{J=0} \approx \frac{G_0^2}{N}\frac{(G_0-D) U(G_0)^2 V(G_0)^2  \left((D-G_0) U(G_0)^2+2 G_0 V(G_0)\right)}{\left(G_0 V(G_0)-(G_0-D) U(G_0)^2\right)^2}
\ee
Here we have eliminated $\lambda^2$ in terms of the self-consistent equation. Let us use this form to prove that $\langle C_{ij}^f C_{ij}^f \rangle_W$ cannot diverge as long as $f(x)$ is bounded by $ x^p$, $p<1$, at large $x$. We already saw that $G_0$ is finite in such a nonlinear network, so this correlation function can diverge only if the denominator vanishes. The terms inside the square in the denominator are
\be
G_0 V(G_0)-(G_0-D) U(G_0)^2 = \langle f(x)^2 \rangle_{G_0} - (G_0-D)  \frac{1}{G_0^2} \,\langle x f(x) \rangle_{G_0}^2\,,
\ee
where $\langle ... \rangle_{G_0}$ denotes the normalized Gaussian average (\ref{eq:Gexp}) but now with respect to a single variable $x$ with variance $G_0$. By the Cauchy-Schwarz inequality,

\be
\langle x f(x) \rangle_{G_0}^2 \le \langle x x \rangle_{G_0}\,\langle f(x) f(x) \rangle_{G_0} = G_0\,\langle f(x) f(x) \rangle_{G_0}\,.
\ee
Therefore,
\be
G_0 V(G_0)-(G_0-D) U(G_0)^2 \ge \langle f(x)^2 \rangle_{G_0}- (G_0-D)  \frac{1}{G_0} \langle f(x)^2 \rangle_{G_0} = \frac{D}{G_0}\, \langle f(x)^2 \rangle_{G_0}\,.
\ee
The right hand side is strictly positive, and we conclude that in the presence of interactions $f(x)$ bounded by $x^p$ with $p<1$ at large $x$, the correlation of off-diagonal components of the covariance matrix is finite at all coupling.

This analysis also yields the statistics of higher correlation functions. For instance, at this order of the large $N$ expansion
\be
\langle f_i f_i f_i f_i \rangle_{\xi,\,W} = \frac{1}{Z_{N_r}} \frac{\partial^2 Z_{N_r}}{ (\partial J_i^{aa})^2} \Big|_{J=0} \approx G_0^2 (W(G_0)+3)\,.
\ee

A general property of the large $N$ expansion is that, once a leading nonzero correlator is identified, higher point correlation functions factorize. Two examples are shown in (\ref{eq:factorize}) and (\ref{eq:factorize2}). Similarly, having identified the leading off-diagonal contributions (\ref{eq:off}), higher-point correlators including it will factorize. For instance,

\be
\langle C_{ij}^f C_{ij}^f C_{kk}^f C_{ll}^f \rangle_W \approx \langle C_{ij}^f C_{ij}^f \rangle_W \langle C_{kk}^f \rangle_W \langle C_{ll}^f \rangle_W\,.
\ee

An important consequence of this property is that we now have a complete characterization of the probability distribution $P(C^f)$ of the covariance matrix $C^f_{ij}$. From Eq.~\eqref{eq:final1} we retain only the terms that generate the leading cumulants relevant for $P(C^f)$: the mean $\langle C^f_{ii}\rangle_W$, the local quadratic contribution, and the leading $O(1/N)$ variance of
off-diagonal entries. The remaining terms in Eq.~\eqref{eq:final1} generate cumulants that are not
needed for $P(C^f)$ at this order (or are fixed by large-$N$ factorization) and are therefore omitted in
$Z_{C^f}(J)$. Finally, we use the self-consistent equation \eqref{eq:Gself} to eliminate $\lambda^2$, which yields
\bea\label{eq:final2}
\log Z_{C^f}(J) &\approx & G_0 V(G_0) \sum_j \sum_a J_j^{aa}+ \frac{1}{2} G_0^2 V(G_0)^2 \sum_j \sum_{a<b} (J_j^{ab})^2 \\
&+&\frac{1}{N}\frac{G_0^2 (G_0-D) U(G_0)^2 V(G_0)^2  \left((D-G_0) U(G_0)^2+2 G_0 V(G_0)\right)}{2 \left(G_0 V(G_0)-(G_0-D) U(G_0)^2\right)^2} \sum_{a<b} (\sum_j J_j^{ab})^2 \nonumber\,.
\eea
 We conclude that in the large $N$ limit, the statistics of $C^f$ is determined by the leading nonzero
correlators \eqref{eq:C2point} and \eqref{eq:off}.

\subsection{Dimension of participation}

Computations in neural systems  usually depend on structured neural activity whose dimensionality is lower than the one of the state space \cite{gao2017theory,jazayeri2021interpreting}, that would be $N$ for our system. One of the most usual ways to quantify dimensionality is the {\it participation dimension}. It is derived from the eigenspectrum of the neuronal covariance matrix. This matrix underlies Principal Component Analysis \cite{jolliffe2005principal},  and indicates how pairs of neurons covary across time and task parameters \cite{gao2017theory}. It is defined by
\begin{equation}
 D_{PR} = \frac{(\sum_{i=1}^N{\lambda_i})^2}{\sum_{i=1}^N{\lambda_i^2}}\,,
\end{equation}
where $\lambda_i$ are the eigenvalues of the covariance matrix.
The motivation for this definition is that if the activity has predominant variability along a set of $D$ directions, then there will be $D$ eigenvalues that are similar between them but much larger than the rest and we will have $D_{PR} \approx D$.

Rewriting this expression in terms of the trace of the covariance matrix and its square we obtain that
\be
D_{PR}^f= \frac{(\sum_i C^f_{ii})^2}{ \sum_{i,j} C^f_{ij} C^f_{ij}}= \frac{N (\langle C_{11}^f  \rangle_W)^2}{ \langle (C^f_{11})^2  \rangle_W+(N-1)\langle (C^f_{12})^2\rangle_W}\,.
\ee
We used the self-averaging property to relate the sum over neurons to the $W$-average, and the homogeneity of the network to fix the sites at $1$ and $2$. At large $N$, and recalling that $\langle (C^f_{11})^2  \rangle_W \approx (\langle C_{11}^f  \rangle_W)^2$, this becomes
\be
D_{PR}^f=N \frac{1}{1+\langle (C^f_{12})^2\rangle_W/(\langle C_{11}^f  \rangle_W)^2}\,.
\ee
Using the results of the previous subsection, we obtain
\be\label{eq:prout}
D_{PR}^f=N \left(1-\lambda^2 U(G_0)^2\right)^2=N \left(1- \frac{G_0-D}{G_0 V(G_0)} U(G_0)^2 \right)^2\,,
\ee
where in the second equality we have eliminated the coupling using the self-consistent equation. By the same arguments as before in terms of the Cauchy-Schwarz inequality, the participation dimension is strictly positive for activation functions bounded by $x^p$ with $p<1$.

All these steps can be redone with minor modifications in order to compute the statistics of the correlation functions of inputs $\phi_i$; see the Appendix \ref{app:input} for more details. In particular, the participation dimension for inputs becomes, from (\ref{eq:prin}),
\be\label{eq:prinmain}
D_{PR}^\phi=N\,\frac{\left(1-\lambda ^2 U(G_0)^2\right)^2}{1- \lambda ^2 U(G_0)^2+(1/2)\lambda ^4 V(G_0)^2}\,.
\ee
Therefore, except in the linear network, the effective dimension for outputs and inputs is different. We will evaluate explicitly these correlation functions for different networks in the next section.


\section{Applications}\label{sec:appl}

In this section we will consider applications of our general results above. We will first consider a class of power-law activation functions for which we obtain explicit analytic expressions for the correlation functions. They are characterized by power-law dependence on the coupling $\lambda$, and are useful for approximating more realistic networks in different regimes, such as near saturation, weak coupling, etc. We next introduce a class of ``Padé activation functions'' that are very useful for simultaneously capturing the small $\phi$ and large $\phi$ regimes of the activation functions. With this choice we will also be able to give explicit results for the correlation functions (albeit in terms of special functions). We will then perform extensive comparisons with numerical results for two activation functions of interest, and we will assess the behavior of $1/N$ corrections and the approach to equilibrium.

\subsection{Power-law activations}\label{subsec:power-law}

As our first application, we will obtain analytic results for activation functions of the form
\be\label{eq:powerf}
f(\phi) = \alpha \, sgn(\phi)|\phi|^p\;,\;0\le p \le 1\,.
\ee
Such power-law activation functions appear in different contexts. The exponent $p=1$ corresponds to the linear activation function; this is also a good approximation for odd activation functions and sufficiently small $\phi$. The case $p \to 0$ applies to a regime near saturation, such as large $\phi$ for the $\tanh(x)$ activation. Other powers, such as $p=1/2$ can be relevant over broad intermediate regimes for $\phi$. In fact $p=1/2$ is obtained for neural dynamics that undergo a type I  bifurcation that leads to spike generation \cite{ermentrout1996type,gutkin1998dynamics,izhikevich2007dynamical}. For our present analysis, we will assume that the neural activation function admits a range of parameters where it can be well approximated by (\ref{eq:powerf}).

The functions $V(G)$ and $U(G)$ introduced in (\ref{eq:Vdef}) and (\ref{eq:Udef}), respectively, evaluate to
\be
V(G)= \frac{ 2^p  \Gamma \left(p+\frac{1}{2}\right)}{\sqrt{\pi }}\alpha^2\,G^{-1+p}\,,\, U(G)=\frac{2^{\frac{p+1}{2}}  \Gamma \left(\frac{p}{2}+1\right)}{\sqrt{\pi }}\alpha  \, G^{-\frac{1-p}{2}}\,.
\ee
The self-consistent equation for the 2-point function $\langle \phi_i \phi_j \rangle_{\xi, W}= G_0$ becomes, from (\ref{eq:Gself}),
\be\label{eq:selfpower}
G_0= D+\frac{ 2^p  \Gamma \left(p+\frac{1}{2}\right)}{\sqrt{\pi }}\alpha^2\, \lambda^2\, G_0^p\,.
\ee
For $p=1$, this reproduces the standard 2-point function for linear networks,
\be
G_0 = \frac{D}{1-\alpha^2 \lambda^2}\,.
\ee
$\alpha^2 \lambda^2 \to 1$ corresponds to the instability of the linear network. This instability is avoided by the interactions created by the nonlinear activation function.

On the other hand, taking $p \to 0$ we see that the 2-point function for networks in the saturation regime is
\be
G_0= D+ \alpha^2 \lambda^2\,.
\ee
For more general $p$, (\ref{eq:selfpower}) can be solved in a power series of $\lambda^2$ for small $\lambda$. This is the weakly coupled limit of the interacting network. The self-consistent equation also allows us to obtain the 2-point function in the limit of strong coupling (large $\lambda$ or small $D$). The result is

\be
G_0 \approx  \left(\frac{ 2^p \alpha^2  \Gamma \left(p+\frac{1}{2}\right)}{\sqrt{\pi }}\right)^{1/(1-p)} \, \lambda^{2/(1-p)}\,.
\ee

The 2-point correlation of the outputs then scales as

\be
C^f_{ij}= \langle f_i f_j \rangle_{W, \xi} \sim \delta_{ij}\lambda^{2p/(1-p)}\,.
\ee

This shows concretely how any sub-linear transfer function is enough to suppress the divergence that appears in linear systems.This is compatible with the result one would obtain for the analysis of a rate model in the mean field limit \cite{treves1993mean}.

Plugging the explicit expressions for $V$ and $U$ in (\ref{eq:off}), we obtain the variance of the off-diagonal elements of the covariance matrix:

\be\label{eq:ratioscaling}
\frac{\langle C_{ij}^f C_{ij}^f \rangle_W}{ \langle C_{ii}^f\rangle_W \langle C_{jj}^f\rangle_W} =\frac{1}{N} \frac{4 (G_0-D) \Gamma \left(\frac{p}{2}+1\right)^2 \left(\sqrt{\pi }  \Gamma \left(p+\frac{1}{2}\right)G_0- \Gamma \left(\frac{p}{2}+1\right)^2(G_0-D)\right)}{\left(\sqrt{\pi }  \Gamma \left(p+\frac{1}{2}\right)G_0-2 \Gamma \left(\frac{p}{2}+1\right)^2 (G_0-D)\right)^2}\,.
\ee

In the three regimes that we discussed above, this expression simplifies to

\begin{align}
\frac{\langle C_{ij}^f C_{ij}^f \rangle_W}{ \langle C_{ii}^f\rangle_W \langle C_{jj}^f\rangle_W} &=\frac{1}{N}\frac{\alpha ^2 \lambda ^2 \left(2-\alpha ^2 \lambda ^2\right)}{\left(1-\alpha ^2 \lambda ^2\right)^2}\,,  \;( \, p=1\;,\;\textrm{linear regime}) \notag \\
\frac{\langle C_{ij}^f C_{ij}^f \rangle_W}{ \langle C_{ii}^f\rangle_W \langle C_{jj}^f\rangle_W} &=\frac{1}{N}\frac{4 \alpha ^2 \lambda ^2 \left((\pi -1) \alpha ^2 \lambda ^2+\pi  D\right)}{\left((\pi -2) \alpha ^2 \lambda ^2+\pi  D\right)^2} \,, \;(\,p=0\;,\;\textrm{saturation})  \\
\frac{\langle C_{ij}^f C_{ij}^f \rangle_W}{ \langle C_{ii}^f\rangle_W \langle C_{jj}^f\rangle_W} &=\frac{1}{N}\frac{4 \Gamma \left(\frac{p}{2}+1\right)^2 \left(-\Gamma \left(\frac{p}{2}+1\right)^2+\sqrt{\pi } \Gamma \left(p+\frac{1}{2}\right)\right)}{\left(\sqrt{\pi } \Gamma \left(p+\frac{1}{2}\right)-2 \Gamma \left(\frac{p}{2}+1\right)^2\right)^2}\,, \;(\,p<1\;,\;\textrm{strong coupling})\,. \notag
\end{align}

We can also evaluate explicitly the participation dimension.
Using (\ref{eq:ratioscaling}), we obtain

\be\label{eq:dprscaling}
D_{PR}^f= N \frac{\left(\sqrt{\pi }  \Gamma \left(p+\frac{1}{2}\right)G_0-2 (G_0-D) \Gamma \left(\frac{p}{2}+1\right)^2\right)^2}{\pi   \Gamma \left(p+\frac{1}{2}\right)^2G_0^2}\,.
\ee

In the three regimes we are discussing, the effective participation dimension evaluates to

\begin{align}
D_{PR}^f &=N \left(1-\alpha ^2 \lambda ^2\right)^2\,,  \;( \,p=1\;,\;\textrm{linear regime}) \notag \\
D_{PR}^f &=N \frac{\left((\pi -2) \alpha ^2 \lambda ^2+\pi  D\right)^2}{\pi ^2 \left(\alpha ^2 \lambda ^2+D\right)^2} \,, \;(\,p=0\;,\;\textrm{saturation})  \\
D_{PR}^f &=N\frac{\left(\sqrt{\pi } \Gamma \left(p+\frac{1}{2}\right)-2 \Gamma \left(\frac{p}{2}+1\right)^2\right)^2}{\pi  \Gamma \left(p+\frac{1}{2}\right)^2}\,, \;(\,p<1\;,\;\textrm{strong coupling})\,. \notag
\end{align}

The result for the participation dimension of the linear network has been obtained before in \cite{dahmen2020strong,doi:10.1073/pnas.1818972116,hu2022spectrum}; the effective dimension vanishes as $\alpha^2 \lambda^2 \to 1$, corresponding to the onset of the instability of the linear network. On the other hand, for nonlinear networks, with $p<1$, this function does not vanish. The effective dimension is monotonically decreasing with $p$, attaining its maximum at $p=0$ (saturation regime), and vanishing for $p=1$ at the instability threshold.

\begin{figure}[htb]
    \centering
    \subfigure{\includegraphics[width=0.32\textwidth]{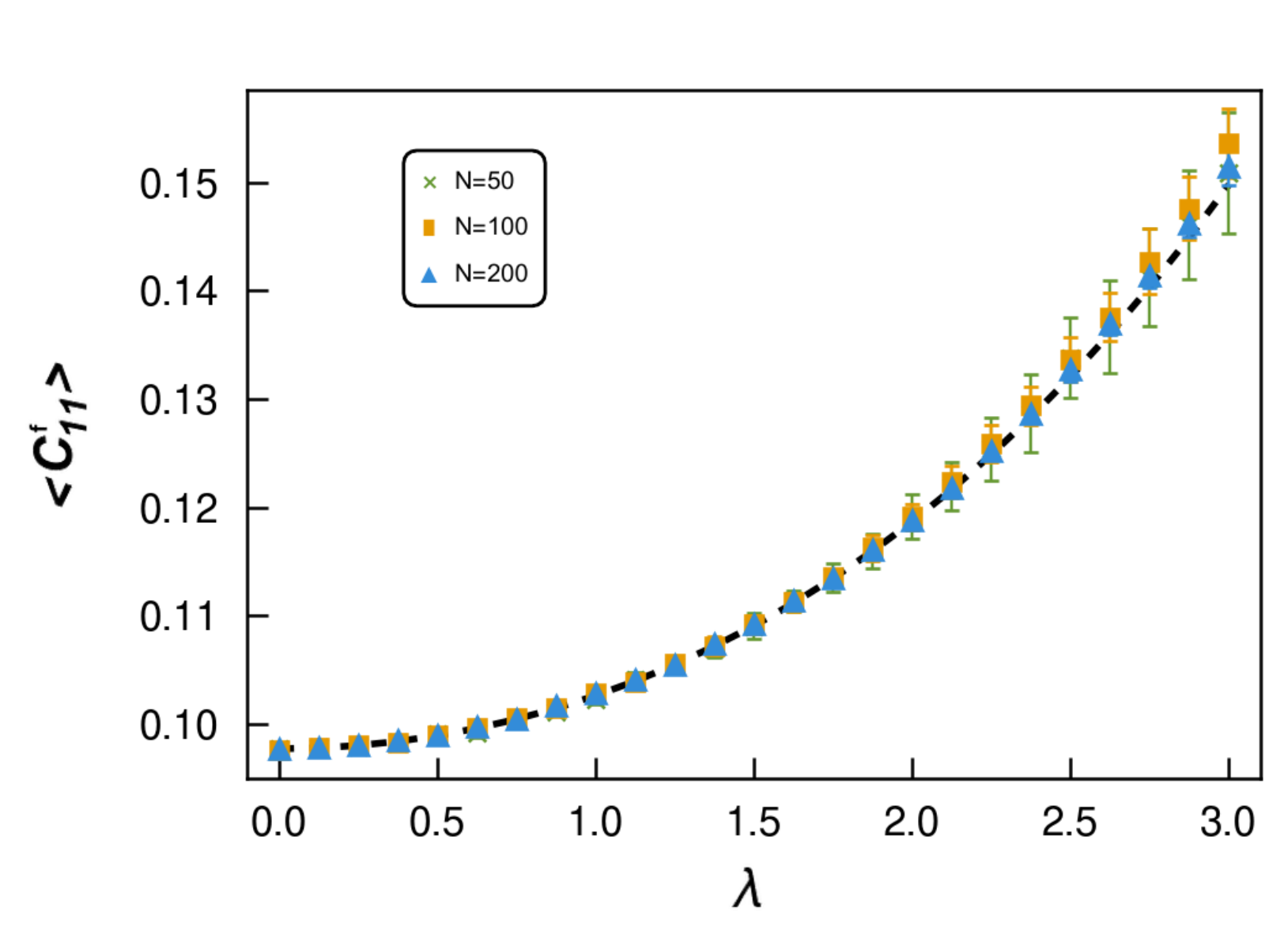}}
    \subfigure{\includegraphics[width=0.34\textwidth]{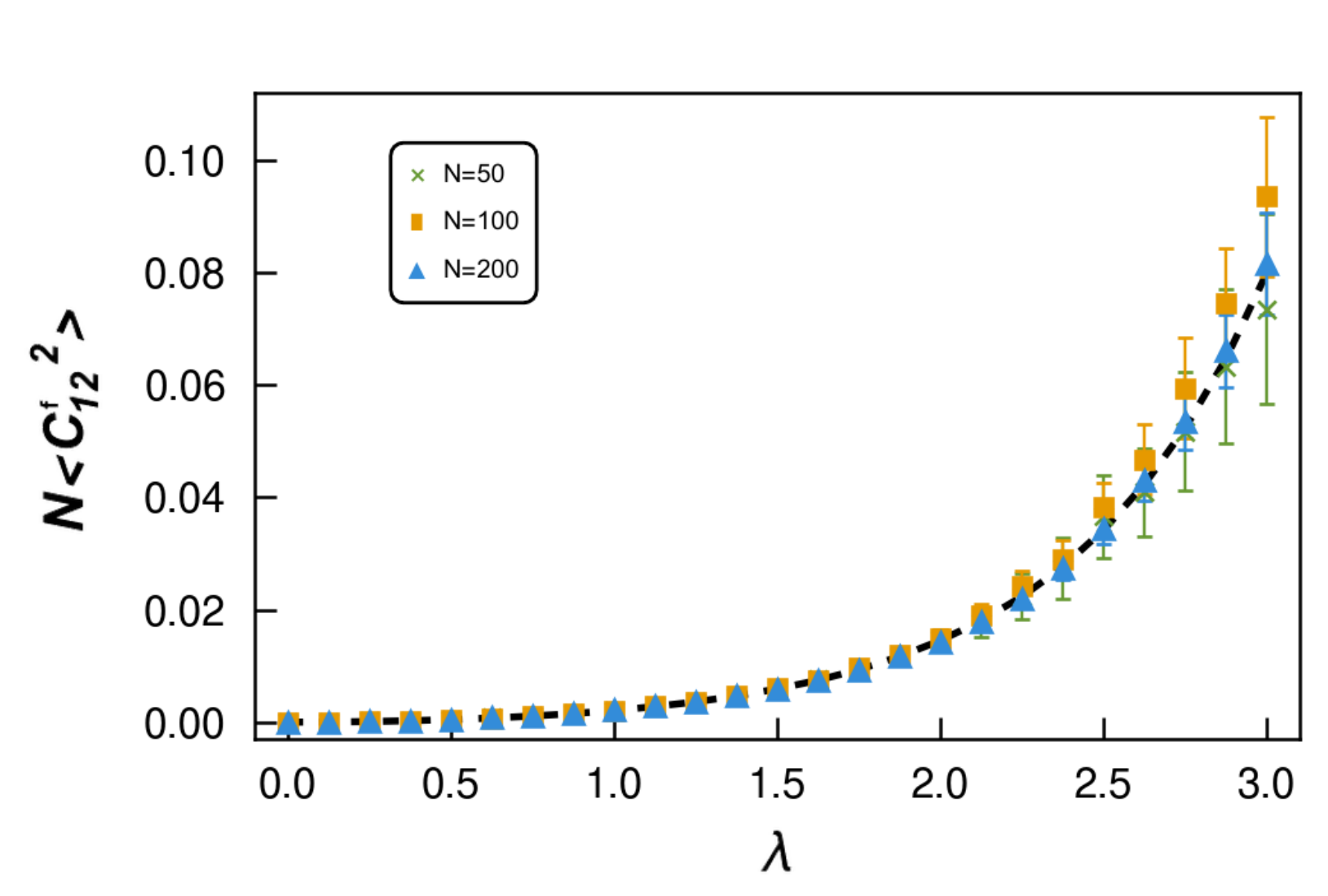}}
    \subfigure{\includegraphics[width=0.32\textwidth]{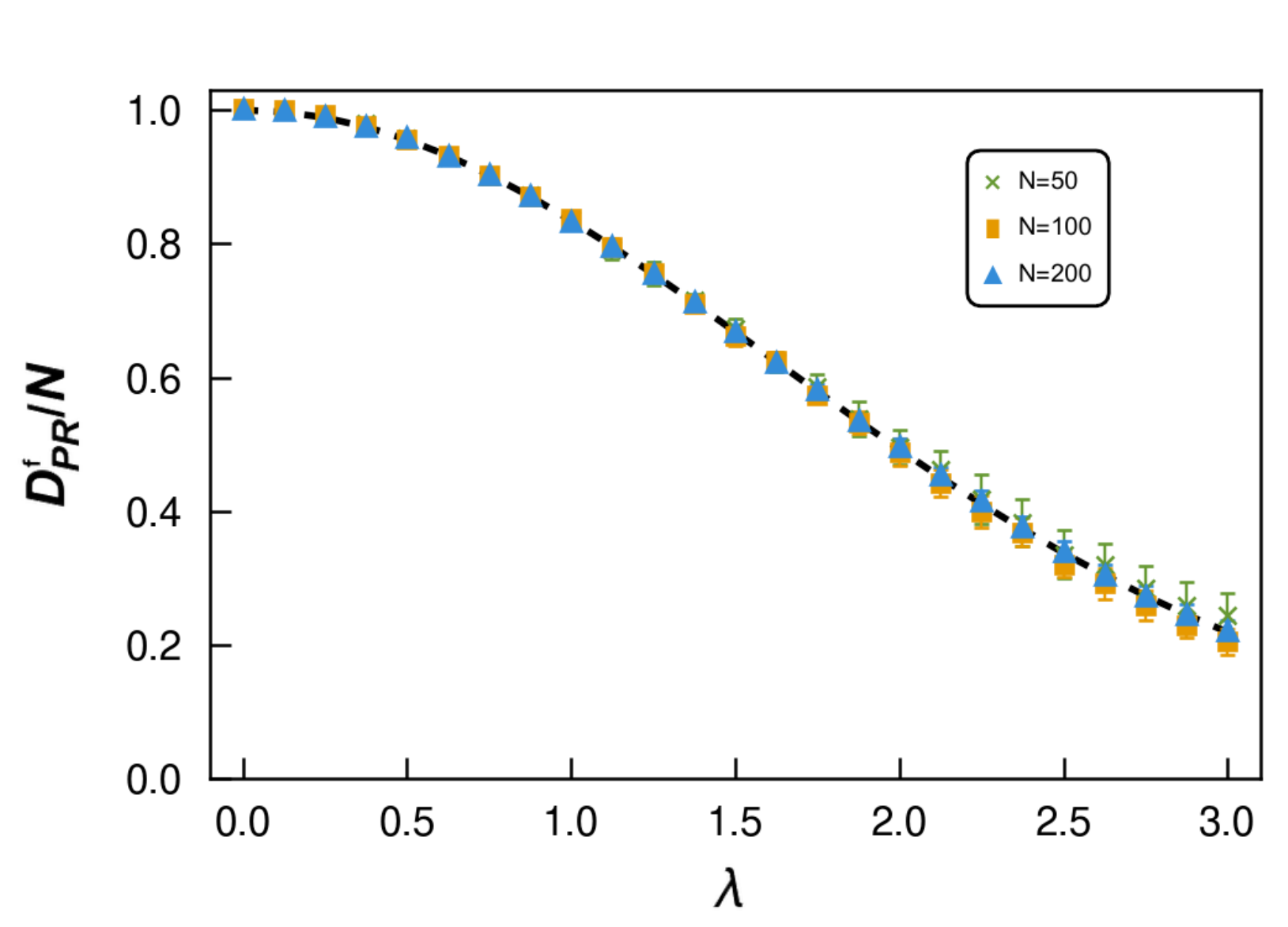}}
    \caption{Comparison of analytical (dashed line) and numerical results (dots) for the output correlations. Left panel: average value of the diagonal correlation. Central panel: average of the square of the off-diagonal correlators. Right panel: participation ratio (participation dimension divided by network size). We use the transfer function of Eq.(\ref{eq:powerf}) with $\alpha=0.35$, $p=1/2$ and $D=1$; the network sizes are $N=50, 100, 200$. The error bars represent the standard deviation over 5 simulations with different realizations of the coupling matrix $W$ and noise $\xi$.}
    \label{fig:cout_power}
\end{figure}

Figure~\ref{fig:cout_power} compares these analytical predictions with
numerical simulations for the power-law activation function with \(p=1/2\), \(D=1\) and \(\alpha=0.35\). The three panels show, respectively, the average values of the diagonal and non-diagonal terms of the covariance matrix and the participation dimension of the outputs. The agreement with the leading large-\(N\) prediction is good over the range of couplings considered.

\subsection{Padé activations and numerics}

In this section we will perform detailed comparisons between our large $N$ formulas and numerical results. The details of the numerical methods are presented in Appendix B.

We introduce a class of nonlinear transfer functions that we  call ``Padé activation functions'':
\be\label{eq:pade1}
f(\phi)=\frac{\phi}{\sqrt{1+\beta^2 (\phi^2)^{1-p}}}\,,
\ee
where the parameter $\beta$ controls the strength of the nonlinear behavior. One motivation for this is that many neural systems can be modeled by an activation function that is linear at small input $\phi$, but exhibits saturation at large input, or a power-law behavior over an intermediate but large regime. The Taylor expansion approximation can in general capture either the limit of small or large $\phi$, but not both.  The idea of Padé approximants is that they can capture both regimes simultaneously by performing an approximation in terms of ratios of polynomials \cite{BakerGravesMorris1996}. This type of activation function has been previously analyzed in the context of convolutional neural networks \cite{kelecs2024paon}
and deep learning \cite{molina2020pade} but not in RNNs, as far as we know.
This expansion allows to perform analytically the Gaussian integrals such as  Eqs.~(\ref{eq:Vdef}) and (\ref{eq:Udef}). This motivated our choice of (\ref{eq:pade1}): its square is a ratio of polynomials with the desired small and large $\phi$ behavior. In particular, for $p=0$ the square of (\ref{eq:pade1})  is the Padé approximant of $(\tanh(\beta \phi)/\beta)^2$. In what follows we will focus on two cases: $p=0$ and $p=1/2$.

\subsubsection{Activation function with $p=0$}

Let us first evaluate our results and compare with numerics for an activation function
\be\label{eq:padep0}
f(\phi)=\frac{\phi}{\sqrt{1+\beta^2 \phi^2}}\,.
\ee
This form, unlike the $\tanh(x)$ function, allows us to evaluate explicitly the functions $V(G)$ and $U(G)$ that enter the neural correlators. We find
\begin{align}
V(G) &= \frac{1}{\beta ^2 G}-\frac{\sqrt{\frac{\pi }{2}} e^{\frac{1}{2 \beta ^2 G}} \text{erfc}\left(\frac{1}{\sqrt{2} \beta  \sqrt{G}}\right)}{\beta ^3 G^{3/2}} \notag\\
U(G)&=\frac{U\left(\frac{1}{2},0,\frac{1}{2 G \beta ^2}\right)}{\sqrt{2} \beta  \sqrt{G}}\,,
\end{align}
where $\text{erfc}(x)$ is the complementary error function, and $U(a,b,z)$ is the confluent hypergeometric function.

We now compare these results with the ones obtained using network simulations (see Appendix \ref{app:numerics} for details). In Fig.~\ref{fig:coutp0} we show the average values of the diagonal and non-diagonal terms of the covariance matrix and the participation dimension of the outputs.

\begin{figure}[htb]
    \centering
    \subfigure{\includegraphics[width=0.32\textwidth]{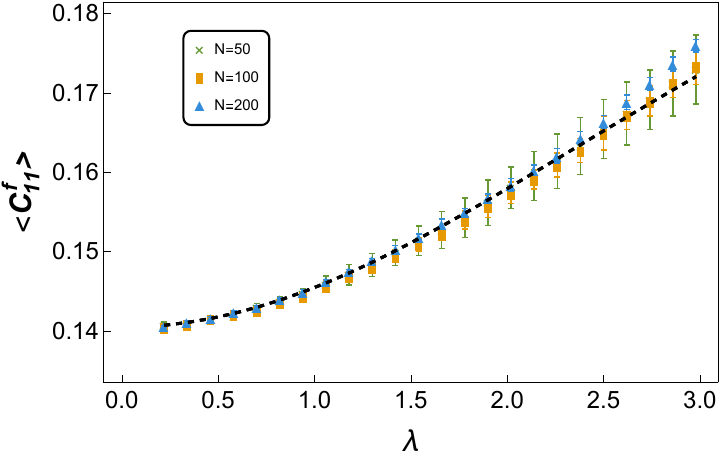}}
    \subfigure{\includegraphics[width=0.32\textwidth]{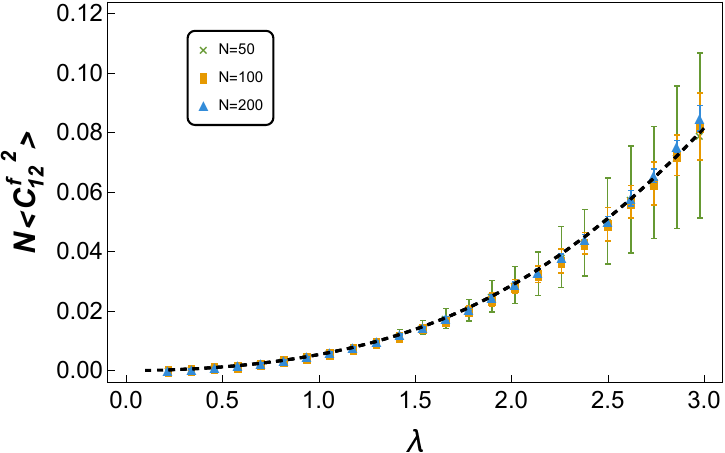}}
    \subfigure{\includegraphics[width=0.32\textwidth]{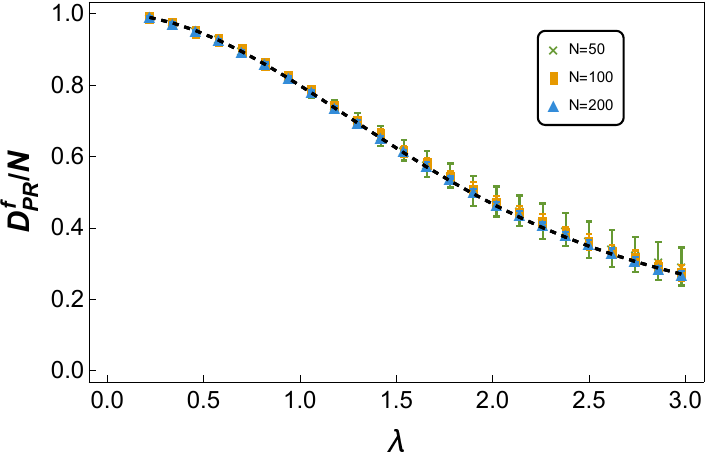}}
    \caption{Comparison of analytical (dashed line) and numerical results (dots) for the output correlations. Left panel: average value of the diagonal correlation. Central panel: average of the square of the off-diagonal correlators. Right panel: participation ratio (participation dimension divided by network size). We use the transfer function of Eq.(~\ref{eq:padep0}) with $\beta=2$ and $D=1$; the network sizes are $N=50, 100, 200$. The error bars represent the standard deviation over 5 simulations with different realizations of the coupling matrix $W$ and noise $\xi$.}
    \label{fig:coutp0}
\end{figure}

\begin{figure}[htb]
    \centering
    \subfigure{\includegraphics[width=0.4\textwidth]{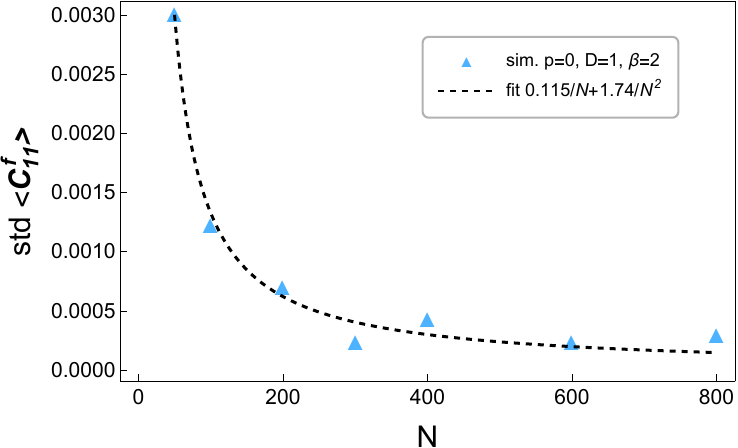}}
    \subfigure{\includegraphics[width=0.4\textwidth]{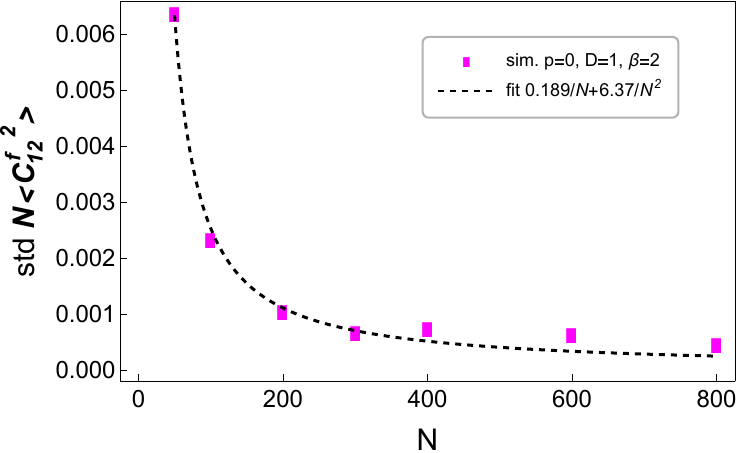}}
    \caption{Scaling with $N$ of the  standard deviations of $<C_{ii}^f>$ (left panel) and $N <C_{ij}^{f 2} >$ (right panel). Dashed line shows a fit in powers of $1/N$: $a/N + b/N^2$.}
    \label{fig:scaln}
\end{figure}

We find an excellent agreement between theory and simulation even for sizes of about a few hundred neurons. In Fig.~\ref{fig:coutp0} it is possible to see that the standard deviation of the different quantities  decrease with   network size. This is compatible with the $1/N$ expansion presented in Section~\ref{subsec:Nlarge}, since the fluctuations are controlled by higher order moments that fall with $N$ with  $1/N^2$ or faster. A key point is that the observables we compare are strongly self-averaging: they involve averages over $\mathcal{O}(N)$ diagonal entries and $\mathcal{O}(N^2)$ off-diagonal pairs (as well as averages over disorder realizations). As a consequence, the standard error decreases rapidly with 
N even before considering additional averaging over disorder realizations.  In order to check this we performed simulations with sizes up to $N=800$ and evaluated the rate of decrease of the standard deviation of the average diagonal and off-diagonal terms of the output correlations. The results are shown on Fig.~\ref{fig:scaln}.

\begin{figure}[h]
    \centering
    \includegraphics[width=0.45\textwidth]{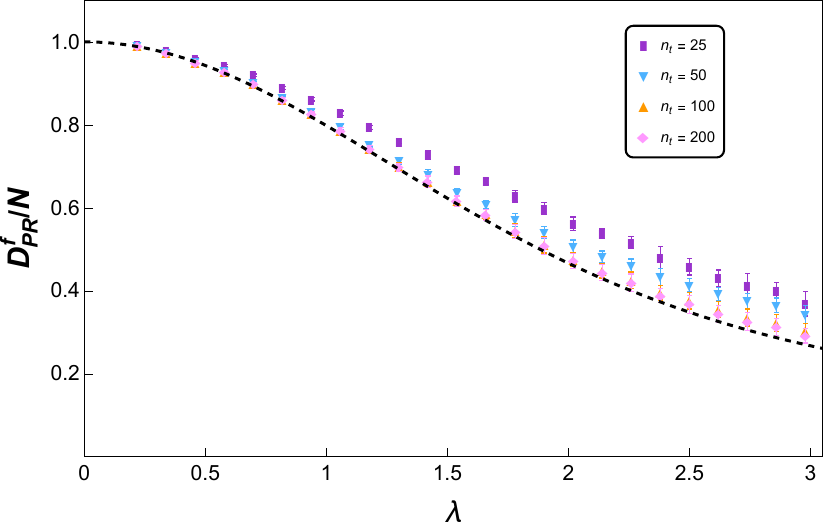}
    \caption{Participation ratio for different simulation times. Dashed line: analytical result. Saturating transfer function ($p=0$), $D=1$ and $\beta=2$. $n_t=100$ correspond to $t_{sim}=1$, or equivalently to $10\tau$}.
    \label{fig:dpritmax}
\end{figure}

In order to test the robustness of the results with respect to the simulation time we also performed simulations with a fixed number of steps
$n_t$, corresponding to a simulation time $t_{sim}=n_t \delta$, without enforcing the convergence condition $d<10^{-4}$. These results are shown in Fig.~\ref{fig:dpritmax}. In these simulations we used $\delta=0.01$ and $\epsilon=0.1$, hence $\tau=0.1$. Therefore $n_t=100$ and $n_t=200$ correspond to $t_{sim}=1$ and $t_{sim}=2$, or
equivalently $10\tau$ and $20\tau$. The convergence observed between
$n_t=100$ and $n_t=200$ is thus consistent with the representative traces
shown in Fig.9, where the quenched dynamics has essentially
reached the fixed point by $t \simeq 2$.

\subsubsection{Activation function with $p=1/2$}

Next we consider,
\be\label{eq:padep1o2}
f(\phi)=\frac{\phi}{\sqrt{1+\beta^2|\phi |}}\,,
\ee
which at large $\phi$ gives a power-law behavior $f\sim |\phi|^{1/2}$. 

The functions $V$ and $U$ are slightly more involved but can still be computed explicitly in terms of special functions,
\begin{align}
V(G)&= -\frac{-2 \sqrt{\pi } F\left(\frac{1}{\sqrt{2} \sqrt{G} \beta ^2}\right)+e^{-\frac{1}{2 \beta ^4 G}} \text{Ei}\left(\frac{1}{2 G \beta ^4}\right)-2 \beta ^4 G+\sqrt{2 \pi } \beta ^2 \sqrt{G}}{\sqrt{2 \pi } \beta ^6 G^{3/2}}\notag \\
U(G)&=\frac{15 \pi  e^{-\frac{1}{4 \beta ^4 G}} \left(\left(1-\beta ^4 G\right) I_{\frac{1}{4}}\left(\frac{1}{4 G \beta ^4}\right)-I_{\frac{3}{4}}\left(\frac{1}{4 G \beta ^4}\right)-I_{\frac{5}{4}}\left(\frac{1}{4 G \beta ^4}\right)+\left(\beta ^4 G+1\right) I_{-\frac{1}{4}}\left(\frac{1}{4 G \beta ^4}\right)\right)}{30 \sqrt{2 \pi } \beta ^6 G^{3/2}}\notag\\
&\qquad- \frac{64 \, _2F_2\left(\frac{3}{2},2;\frac{7}{4},\frac{9}{4};-\frac{1}{2 G \beta ^4}\right)}{30 \sqrt{2 \pi } \beta ^6 G^{3/2}}\,.
\end{align}
Here $F(z)$ is the Dawson integral, $\text{Ei}(z)$ is the exponential integral function, $I_\nu(z)$ is the modified Bessel function of the first kind, and $_{p}F_q(a;b;z)$ is the generalized hypergeometric function.

\begin{figure}[h]
    \centering
    \subfigure{\includegraphics[width=0.32\textwidth]{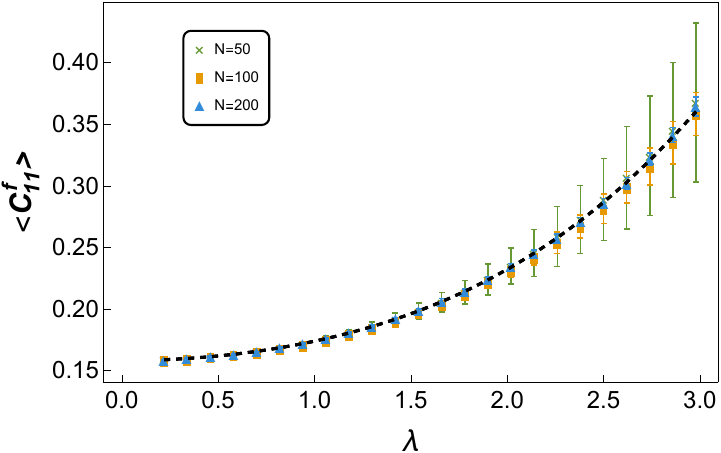}}
    \subfigure{\includegraphics[width=0.32\textwidth]{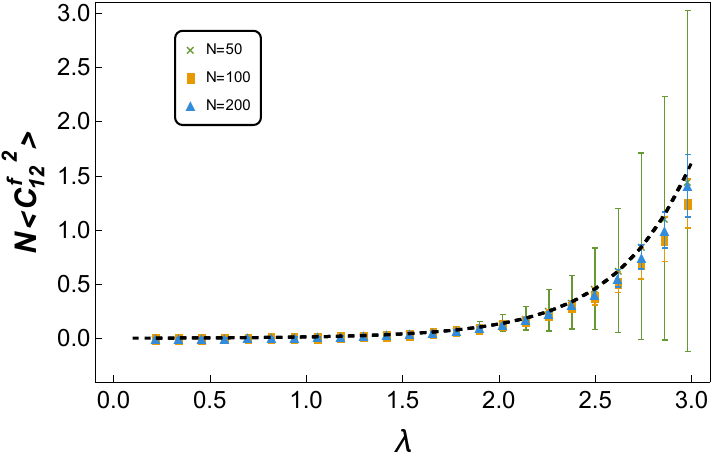}}
    \subfigure{\includegraphics[width=0.32\textwidth]{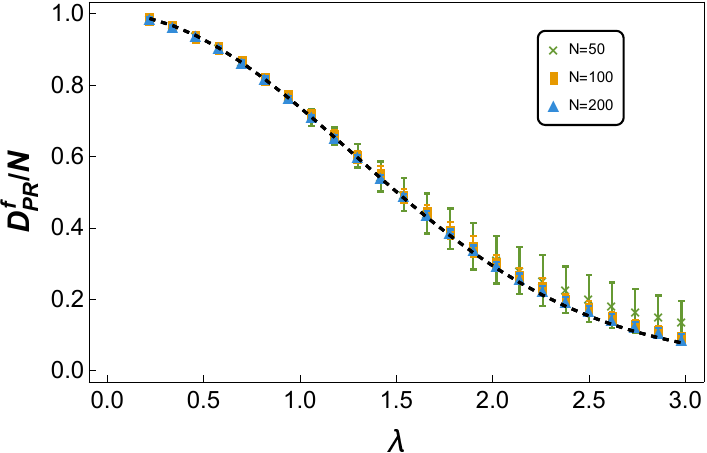}}
    \caption{Comparison of analytical (dashed line) and numerical results (dots) for the output correlation, for the Padé activation function with $p=1/2$. Left panel: average value of the diagonal correlation. Central panel: average off-diagonal correlators. Right panel: participation ratio (participation dimension divided by network size). We use the transfer function of Eq.~\ref{eq:padep1o2} with $\beta=2$ and $D=1$. The network sizes are $N=50, 100, 200$. The error bars represent the standard deviation over 5 simulations with different realizations of the coupling matrix $W$ and noise $\xi$.}
    \label{fig:coutp5}
\end{figure}

In Fig.~\ref{fig:coutp5} we show the results for the non-saturating Pade function of Eq.~(\ref{eq:padep1o2}). We again obtain an excellent agreement between the analytic formulas and the numerical results.

  In the next section we present a brief discussion of related works and more specifically of the difference between our approch and the one in \cite{shen2025covariance}.

\subsection{Comparison to other works and annealed vs quenched disorder}\label{subsec:comparison}

Closely related references to our work include \cite{PhysRevX.8.041029, doi:10.1073/pnas.1818972116}, which also used a path integral representation and the large $N$ limit. Ref. \cite{PhysRevX.8.041029} considered the nonlinear case and focused on the auto-covariance and time-dependent effects, but did not include $1/N$ effects. They obtained a self-consistent equation which plays the role of our (\ref{eq:Gself}). Similar self-consistent equations also appear in other references such as \cite{PhysRevLett.61.259, clark2023dimension}.

On the other hand, Ref. \cite{doi:10.1073/pnas.1818972116} focused on the linear network. In this work a path-integral formulation is also used. In contrast to our work there are no replicas and a linear source term is used. Correlations up to order four are evaluated taking succesive derivatives with respect to the source and the mean values and variances of the elements of the  covariance matrix are determined using a Wick decomposition of the temporal averages. This is possible in the linear case because the statistics for fixed couplings is Gaussian. This approach is not possible in the non-linear case, requiring the use of replicas.  Replicas have also been used in the linear model e.g. in \cite{Layer_2024}.

Let us now analyze in more detail the difference between the quenched regime we have studied in this work, corresponding to $\tau/\tau_\textrm{noise} \ll 1$, and the annealed or white noise limit $\tau_\textrm{noise}/\tau \ll 1$, which has been studied in previous works, such as \cite{doi:10.1073/pnas.1818972116, hu2022spectrum}. In particular,
the recent preprint \cite{shen2025covariance} proposed an ansatz to solve the nonlinear network in terms of an effective coupling in the annealed limit. This was motivated by mean field theory considerations and verified numerically and using random matrix theory.  Their ansatz can be derived explicitly from our large $N$ treatment. For this, we note that eq. (\ref{eq:prout}) motivates the introduction of an effective coupling
\be
\lambda_{eff} \equiv \lambda U(G_0)\,.
\ee
To verify that this gives the proposal of \cite{shen2025covariance}, we observe that integrating by parts in Eq. (\ref{eq:Udef}), we obtain
\be\label{eq:Udefn}
\lambda_{eff} =\lambda \int \frac{dx}{\sqrt{2\pi G}}\,f'(x)\,e^{- \frac{x^2}{2G}}= \lambda \langle f'(x) \rangle_G\,.
\ee
This agrees with the effective connection strength Eq. (3.2) of \cite{shen2025covariance}.

As a further check of the consistency between both formalisms, we note that the output dimension of participation (\ref{eq:prout}) becomes
\be
D_{PR}^f= N (1- \lambda^2 \langle f'(x) \rangle_G^2)^2\,,
\ee
This coincides with Eq. (3.1) of \cite{shen2025covariance} at vanishing frequency.  

The main difference between the quenched and annealed limit lies in the time dependence of the neural correlators and their dependence on the model parameters. In the quenched limit and after reaching equilibrium, the neural correlators in equilibrium are time independent. Specifically, recalling the 2-point functions (\ref{eq:Cphidef}), we have that
\be
\langle C^\phi(t) \rangle_W = \langle \phi_i(t+t_0) \phi_i(t_0) \rangle_{\xi, \,W}\;,\;\langle C^f(t) \rangle_W = \langle f_i(\phi(t+t_0)) f_i(\phi(t_0)) \rangle_{\xi, \,W}\,,
\ee
are independent of $t$ and $t_0$, and coincide with what we have been calling $\langle C^\phi \rangle_W=G$ and $\langle C^f \rangle_W = G V(G)$ without time arguments. This implies that the frequency space correlators have only support at zero frequency:
\be
\langle C^\phi(\omega) \rangle_W = \int dt\,e^{-i \omega t}\,\langle C^\phi(t) \rangle_W= 2\pi \delta(\omega)\,\langle C^\phi(t=0) \rangle_W
\ee
and similarly for $C^f$. For completeness, let us also reproduce here the self-consistent equation (\ref{eq:selfCphi}),
\be\label{eq:selfCphi2}
\langle C^\phi \rangle_W= D_q + \lambda^2 \,\langle C^f \rangle_W\,,
\ee
where, for the following comparison with the annealed limit, we have renamed $D=D_q$ to highlight the quenched limit, and recall that
\be
\langle \xi_i(t+t_0) \xi_j(t_0) \rangle= \delta_{ij} D_q\,,
\ee
independent of time.

In the annealed limit, the self-consistent equation for the frequency space neural correlators is, from Eq. (2.2) of \cite{shen2025covariance},
\be\label{eq:selfCphi3}
(1+\omega^2)\langle C^\phi(\omega) \rangle_W= D_w + \lambda^2 \,\langle C^f(\omega) \rangle_W\,,
\ee
where the white noise coupling $D_w=\sigma^2$ (in their notation) satisfies
\be
\langle \xi_i(t+t_0) \xi_j(t_0) \rangle= \delta_{ij} \delta(t) D_w\,.
\ee

The quenched and annealed noise parameters $D_q$ and $D_w$ cannot be compared directly, as they have different dimensions. However, we can obtain a more precise relation between (\ref{eq:selfCphi2}) and (\ref{eq:selfCphi3}) if instead of $D_w$ we write
\be
D_w= \int dt\,e^{-i \omega t} \,\langle \xi_i(t) \xi_i(0) \rangle = \langle C^\xi(\omega) \rangle\,,
\ee
and propose a self-consistent equation
\be\label{eq:selfCphi4}
(1+\omega^2)\langle C^\phi(\omega) \rangle_W=  \langle C^\xi(\omega) \rangle + \lambda^2 \,\langle C^f(\omega) \rangle_W\,.
\ee
We propose that this self-consistent equation is valid for a more general internal ``colored noise'', corresponding to $\tau_\textrm{noise}$ not parametrically different from $\tau$. We hope to check the consequences of this in future work. It correctly interpolates between the white and quenched limits.  In the white noise limit, (\ref{eq:selfCphi4}) reduces to (\ref{eq:selfCphi3}), while in the quenched limit it reduces to (\ref{eq:selfCphi2}) since all correlators are proportional to $\delta(\omega)$.

\begin{figure}[h]
    \centering
    {\includegraphics[width=0.45\textwidth]{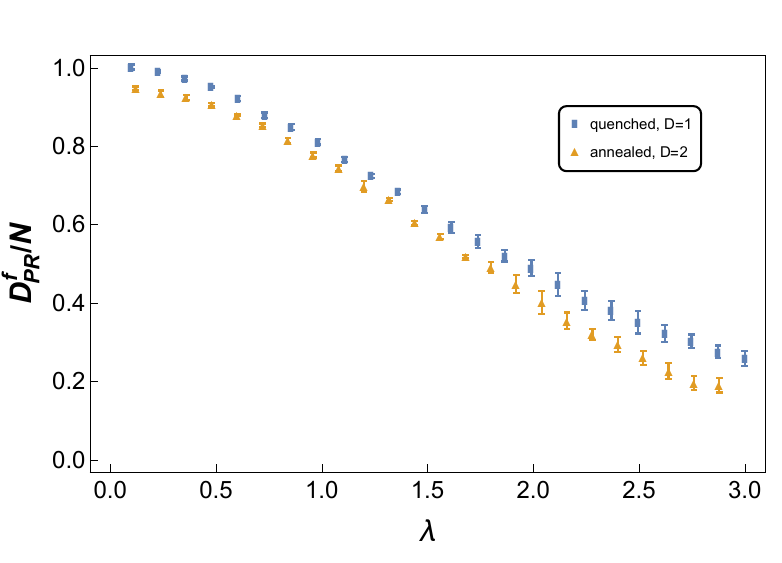}}
    \caption{Comparison of annealed and quenched dynamics with $\tau=1$ for the output correlation for the Padé activation function with $p=1/2$, $N=100, \beta=2$ and $D=2 (1)$ for the annealed (quenched) case.}
    \label{fig:annvsquen}
\end{figure}

To have a better understanding of the theoretical limits and the parameter ranges, let us now compare numerical results in the quenched and white noise limits. In terms of the neural time constant $\tau$, it is natural to contrast results for $D_w \sim D_q/\tau$, which have the same time dimensions. As a concrete example, let us choose
\begin{equation}
D_w=\frac{D_q}{2\tau},
\label{eq:DD}
\end{equation}
which corresponds to a situation where the equal time neural 2-point function in a linear system with white noise agrees with the quenched equilibrium 2-point function.\footnote{Again, this is just an example and there is no fundamental reason for such a relation between $D_w$ and $D_q$, especially in nonlinear cases.}
In Fig \ref{fig:annvsquen} we show the numerical results for $N=100, \tau=1,D_w=2$ and  $D_q=1$ with other parameters as in Fig. \ref{fig:coutp5}. This shows a qualitative similarity between both results in the nonlinear case. We see that even though the quenched and white noise models are quite different mathematical idealizations, they provide complementary results which may not be that different from more experimentally realistic models.

\section{Conclusions}\label{sec:concl}

In this work we have developed a path-integral framework to compute the statistics of correlations in nonlinear recurrent networks, including $1/N$ effects that are required to evaluate cross-correlations of the covariance matrix and the participaton dimension. Our approach generalizes previous treatments of linear networks by incorporating  nonlinear activation functions as interaction terms in the effective action.\footnote {For simplicity we restricted to odd activation functions, which lead to a vanishing neural 1-point function. More general activation functions can be included by allowing for such an expectation value.}  This resolves the instability of the linear theory, yielding strictly positive participation dimensions and a rich set of correlation structures.

A central outcome of our analysis is the emergence of a small set of collective variables that capture the full statistics of correlations.  This is a basic result in large $N$ and disordered average field theories, as well as in the case of linear networks; here we have extended the collective representation to nonlinear networks. This representation makes explicit the way in which global network dynamics constrain local pairwise correlations. It also highlights the universality of some of the results: for instance, the $1/N$ suppression of cross-correlations persists in the nonlinear regime, but their relative fluctuations become essential in controlling the participation dimension. This is consistent with previous findings in the linear case \cite{doi:10.1073/pnas.1818972116,hu2022spectrum}, while substantially extending their range of applicability. We also obtained the generating function of connected neural correlators, focusing for the most part on the covariance matrix; see Sec. \ref{subsec:saddle}. 

The explicit results obtained for power-law and the new class of Padé activation functions illustrate two complementary aspects of the theory. Power-law activations exhibit scaling behavior controlled by the strength of the recurrent coupling, revealing how changes in network connectivity can reshape correlation statistics. Padé activations, on the other hand, provide a flexible family of transfer functions that remain tractable analytically while capturing important features of realistic nonlinearities. The agreement between our analytic predictions and numerical simulations across these cases gives strong support to the validity of the approach. Numerical simulations indicate that the results of the large $N$ limit are relevant even for sizes of a few hundred neurons, while the time required to perform the averages is not required to be larger than about 20 synaptic time constants.

Our results have been derived in the quenched disordered limit, and we have compared to alternative frameworks that use white noise (annealed internal disorder). Dynamical mean-field theory and random matrix methods, which have been successfully applied to related questions \cite{clark2023dimension, zhang2011nonlinear, shen2025covariance}, have provided valuable insights into the statistics of covariance matrices. The path-integral formulation complements these approaches by emphasizing the role of collective variables, the effects of external sources, and the description in terms of the generating function, which naturally encode higher-order correlators and enable systematic expansion in $1/N$. Both the quenched and annealed cases provide mathematically tractable regimes, and can lead to comparable predictions. This suggests that more realistic colored noise models may also behave similarly. Based on the comparison between
these two limits, we have proposed a new self-consistent equation (\ref{eq:selfCphi4}) for the generic colored noise case. We emphasize that
this equation should be regarded as a hypothesis. In future work we plan to derive this from a large $N$ treatment, and study its dynamical consequences.

More broadly, the ability to derive analytic predictions for correlation statistics in nonlinear recurrent networks opens the door to several applications. In neuroscience, it provides a theoretical tool to interpret experimental measurements of correlations, dimensionality, and variability in cortical recordings. In machine learning, the connection between correlation structure and participation dimension may shed light on representational capacity in large recurrent architectures. Finally, from a methodological standpoint, the path-integral formalism can be adapted to other complex systems where nonlinear interactions and collective dynamics shape correlation statistics.

In future work, it will be important to extend our approach to characterize out-of-equilibrium correlation functions, where temporal structure and nonstationary dynamics are expected to play a crucial role. Furthermore, the path-integral formulation provides an efficient framework to evaluate subleading $1/N$ effects using standard diagrammatic methods. For the recurrent networks studied here, such corrections originate from Gaussian fluctuations around the nontrivial saddle point as well as from interaction vertices induced by nonlinear activation functions.

It is interesting that in the quenched regime that our work has addressed, the large $N$ limit seems to work extremely well already for moderate values $N \sim 100$, while the annealed case requires much larger values of $N$.\footnote{We thank an anonymous referee for stressing this.} We have performed a numerical study of the $1/N$ convergence, but it would be interesting to have a detailed analytic understanding of the large $N$ regime of validity. This will require going beyond 4-point functions as well as including ``fluctuation corrections'' from perturbations around the saddle points. Finally, the path-integral framework is also promising for incorporating additional structures, such as plasticity. Exploring these contributions will deepen the connection between neural dynamics and field-theoretic techniques, and we expect to report on these developments in future work.

\section*{Acknowledgments} 
GM is supported by CONICET (PIP grant 112202101 00114CO), CNEA and Instituto Balseiro, Universidad Nacional de Cuyo,
FR is supported by a CONICET PhD fellowship and by Instituto Balseiro, Universidad Nacional de Cuyo.
GT is supported by CONICET (PIP grant 11220200101008CO), CNEA, Instituto Balseiro, Universidad Nacional de Cuyo, and a Simons Foundation targeted grant to institutions.

\appendix
\newpage
\section{Correlations of neural inputs}\label{app:input}

In this Appendix we discuss briefly how to obtain correlations of neural inputs,
\be
 \langle \langle \phi_{i_1} \phi_{i_2} \ldots \rangle_\xi  \langle \phi_{j_1} \phi_{j_2} \ldots \rangle_\xi \ldots \langle \phi_{s_1} \phi_{s_2} \ldots \rangle_\xi  \rangle_W\,,
\ee
This requires modifying (\ref{eq:Zxi}) so that the source couples to $\phi_i \phi_j$ instead of $f_i f_j$. The steps are the same as in Secs. \ref{sec:setup} and \ref{sec:largeN}, with the replacement $J^{ab}_i f^a_i f^b_i \to J^{ab}_i \phi^a_i \phi^b_i$. The net effect is that (\ref{eq:Znr2}) is modified to
\begin{align}
\label{eq:Znrphi}
Z_{N_r}&= \int {\mc D}\phi\, {\mc D}\eta\, {\mc D}\rho\;
\exp\!\left[-\frac{i}{2}\sum_{a,b=1}^{N_r}\eta_{ab}\Big(\sum_{j=1}^{N} g_{ab}(\phi_j)- N \rho_{ab}\Big)\right] \nonumber \\
&\times [\textrm{det}(2\pi M^{-1})]^{N/2}\,
\exp\!\left[-\frac{1}{2}\sum_{j=1}^{N}\sum_{a,b=1}^{N_r}(G^{-1})_{ab}\phi_j^a \phi_j^b\right]\,
\exp\!\left[\sum_{j=1}^{N}\sum_{a,b=1}^{N_r} J^{ab}_j \phi_j^a \phi_j^b\right]. 
\end{align}

The large $N$ effective action (\ref{eq:Seff2}) is then replaced by
\begin{align}\label{eq:Seffphi}
S_{eff}&\approx \frac{1}{2} \log \det(1+i M \eta)-\frac{i}{2}\tr( \eta \rho)+ \frac{i}{2} \tr\left(\eta \langle g \rangle_G\right)+\frac{1}{2} \sum_{a<b}\eta_{ab}^2\langle g_{ab}^2 \rangle_G  \nonumber\\
&-\frac{1}{N}\sum_a (\sum_j J_j^{aa}) G_{aa} -\frac{1}{N} \sum_{a<b}(\sum_j J_j^{ab})\left( G_{ab} - i \eta_{ab} \langle g_{ab}x^a x^b \rangle_G\right)\\
&-\frac{1}{2N} \sum_a \sum_j (J_j^{aa})^2 \left(\langle(x^a x^a)^2 \rangle_G-\langle x^a x^a \rangle_G^2 \right) -\frac{1}{2N}\sum_{a<b} \sum_j (J_j^{ab})^2 \langle ( x^a x^b)^2 \rangle_G\nonumber\,.
\end{align}
Recalling the parametrization of Sec. \ref{subsec:nonlinear}, this becomes
\begin{align}\label{eq:Seffphi3}
S_{eff}&\approx \frac{1}{2} \log \det(1+i M \eta)-\frac{i}{2}\sum_a \eta_{aa} \left[\rho_{aa}- G_{aa} \left(V(G_{aa})-1 \right)\right]\nonumber\\
&-i\sum_{a<b} \eta_{ab} \left(\rho_{ab}- G_{ab} \left[U(G_{aa}) U(G_{bb})-1 \right]\right)\\
&+\frac{1}{2} \sum_{a<b}\eta_{ab}^2 G_{aa} G_{bb}\left[V(G_{aa}) V(G_{bb})-2 U(G_{aa}) U(G_{bb})+1 \right] -\frac{1}{N}\sum_a (\sum_j J_j^{aa})  G_{aa} \nonumber \\
&-\frac{1}{N} \sum_{a<b}(\sum_j J_j^{ab})\left( G_{ab} - i \eta_{ab} G_{aa} G_{bb}\left[U(G_{aa})U(G_{bb})-1 \right]\right)\nonumber\\
&-\frac{1}{2N}\sum_j \left [2 \sum_a  (J_j^{aa})^2\, G_{aa}^2 +\sum_{a<b} (J_j^{ab})^2\,G_{aa} G_{bb} \right ]\nonumber\,.
\end{align}

Computing the saddle point for $\rho$ and $\eta$ in a $1/N$ expansion and plugging back into the effective action, we arrive at the generating function for connected correlators,
\begin{align}\label{eq:finalphi}
\log Z_{N_r}(J) &\approx  G_0  \sum_j \sum_a J_j^{aa}+ \frac{1}{2} G_0^2  \sum_j \sum_{a<b} (J_j^{ab})^2 +G_0^2  \sum_j \sum_a (J_j^{aa})^2\\
&+\frac{1}{N} \frac{G_0^2 (G_0-D) (D+2 G_0 V-G_0) }{\left(G_0 (D-G_0) V'+D V\right)^2} \sum_a (\sum_j J_j^{aa})^2 \nonumber\\
&+\frac{1}{N}\frac{G_0^2 (G_0-D)  \left(2(D-G_0) U^4+2 G_0 U^2V+(G_0-D) V^2\right)}{2 \left((D-G_0) U^2+G_0 V\right)^2} \sum_{a<b} (\sum_j J_j^{ab})^2 \nonumber\,,
\end{align}
which should be compared with the analog expression (\ref{eq:final1}) for the outputs. 

Here $U=U(G_0)$ and $V=V(G_0)$ are the functions defined in the main text, and $V'=V'(G_0)$.

At leading order in the large $N$ expansion, the two-point function is
\be
\langle \phi_i \phi_j \rangle_{\xi, W}= \langle C_{ij}^\phi \rangle_W = \delta_{ij} G_0\,,
\ee
and $G_0$ satisfies the self-consistent equation (\ref{eq:Gself}),
\be
G_0=D+G_0 \,\lambda ^2\, V(G_0)\,.
\ee
Diagonal correlators of the $\phi$ covariance matrix factorize, 
\be\label{eq:factorizephi}
\langle C_{ii}^\phi C_{jj}^\phi \rangle_W \approx \langle C_{ii}^\phi\rangle_W \,\langle C_{jj}^\phi\rangle_W\,,
\ee 
while the off-diagonal correlators are given by
\be
\langle C_{12} C_{12} \rangle_W= \frac{1}{N}\frac{G_0^2 (G_0-D)  \left(2(D-G_0) U^4+2 G_0 U^2V+(G_0-D) V^2\right)}{\left((D-G_0) U^2+G_0 V\right)^2}\,.
\ee
This predicts a participation dimension for inputs
\be\label{eq:prin}
D_{PR}^\phi=N\frac{\left((D-G_0) U^2+G_0 V\right)^2}{\left(D^2-2 D G_0+2 G_0^2\right) V^2-(G_0-D)^2 U^4}\,.
\ee

We have written all these expressions by eliminating $\lambda^2$ in terms of the self-consistent equation. In the main text, we give an equivalent expression for the participation dimension in terms of $\lambda^2$, see (\ref{eq:prinmain}).

\begin{figure}[h]
    \centering
     \includegraphics[width=0.35\textwidth]{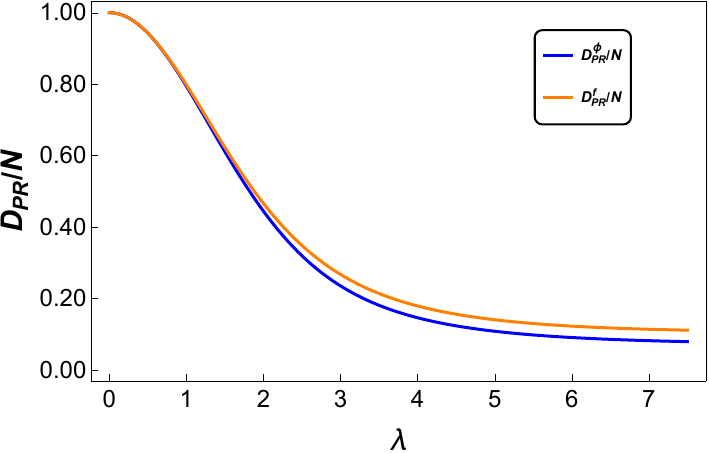}
        \includegraphics[width=0.35\textwidth]{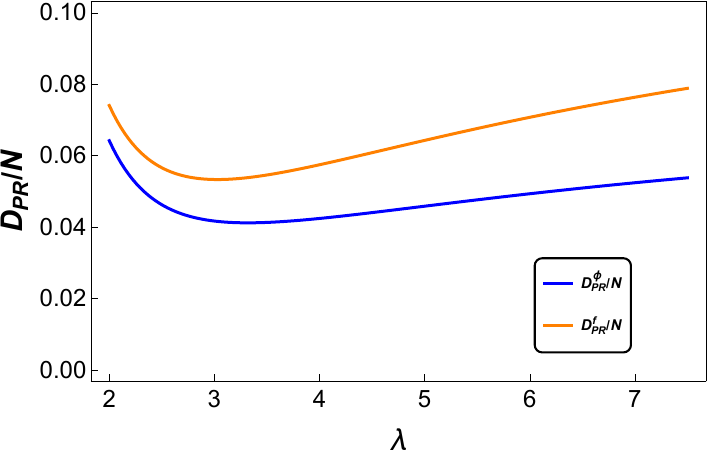}
    \caption{Comparison of participation ratios of inputs and outputs, for $\beta=2, D=1$ (left) and $\beta=2, D=0.1$ (right), with a Pade activation function $p=0$ (\ref{eq:padep0}). Note the difference in scales in the y-axis between the plots.}
    \label{fig:Dprp0}
\end{figure}

In Fig.~\ref{fig:Dprp0}  we show the comparison between input and output participation ratios. We see two different behaviors: for the high noise case $D>1$, we find that $D_{PR}$ decreases with $\lambda$ and the gap between the input and output dimensions is very small. In contrast, in the case of low noise we find an increasing $D_{PR}$ and a significant dimension gap, in agreement with \cite{clark2023dimension}.

\begin{figure}[h]
    \centering
    \subfigure{\includegraphics[width=0.33\textwidth]{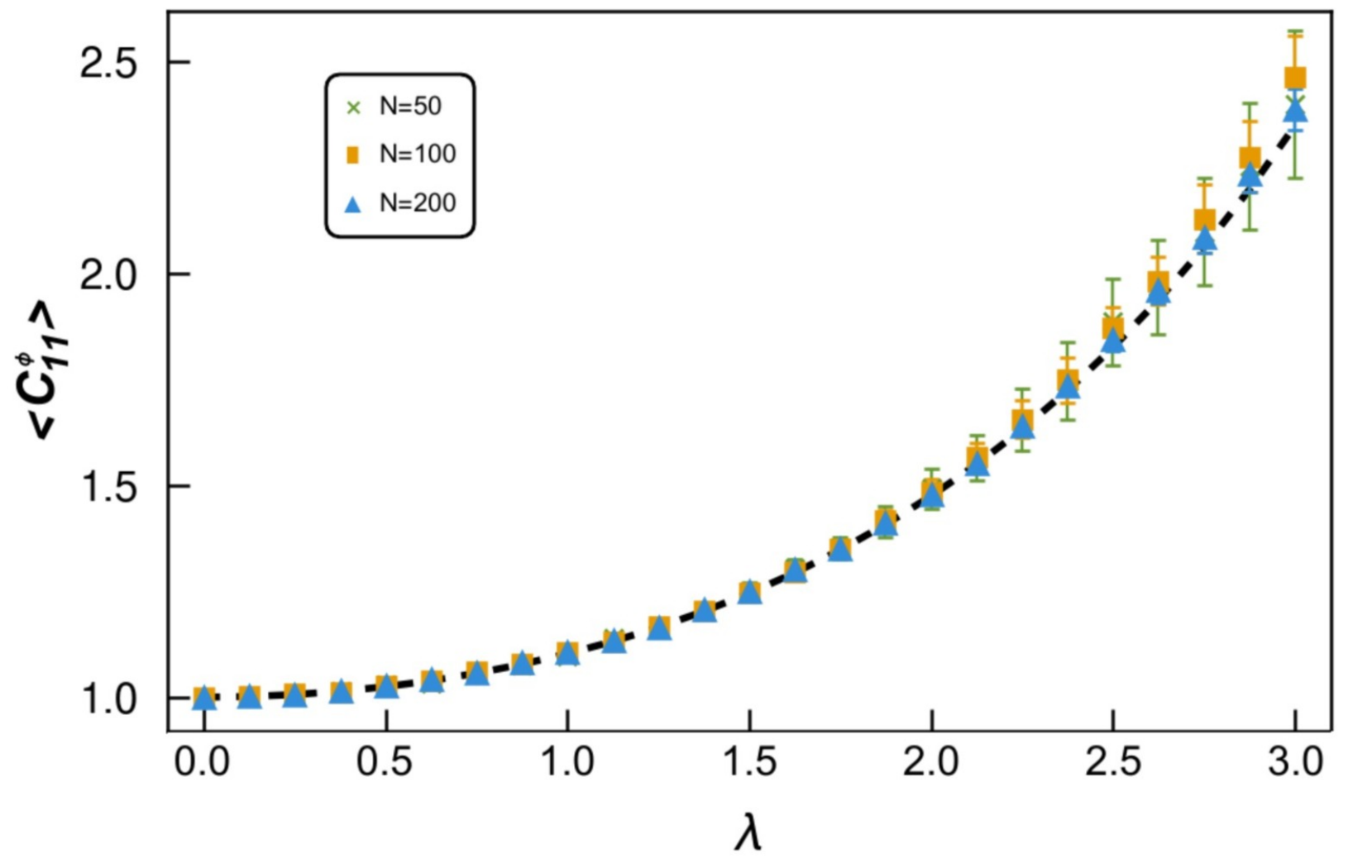}}\hfill
    \subfigure{\includegraphics[width=0.33\textwidth]{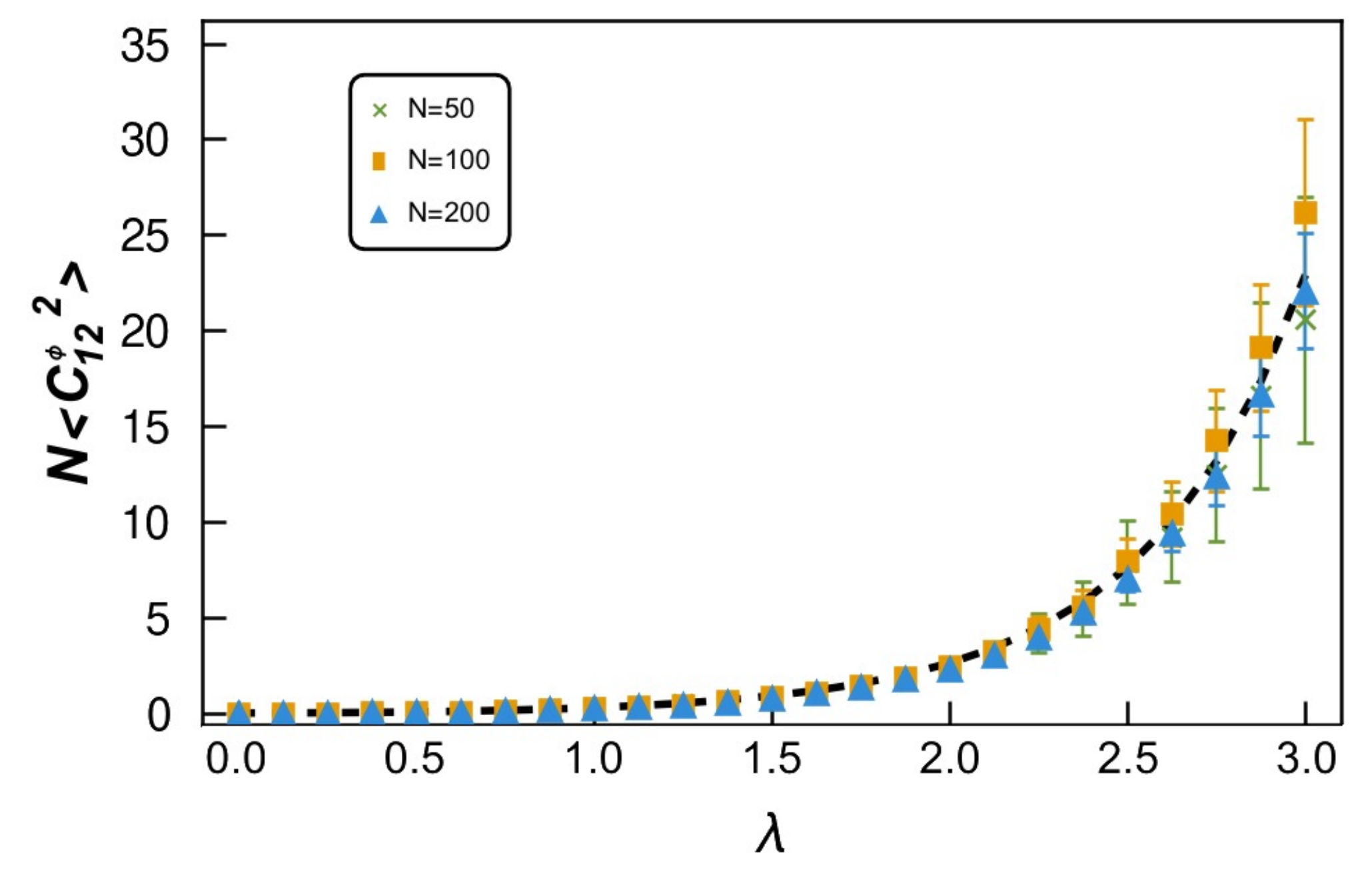}}\hfill
    \subfigure{\includegraphics[width=0.31\textwidth]{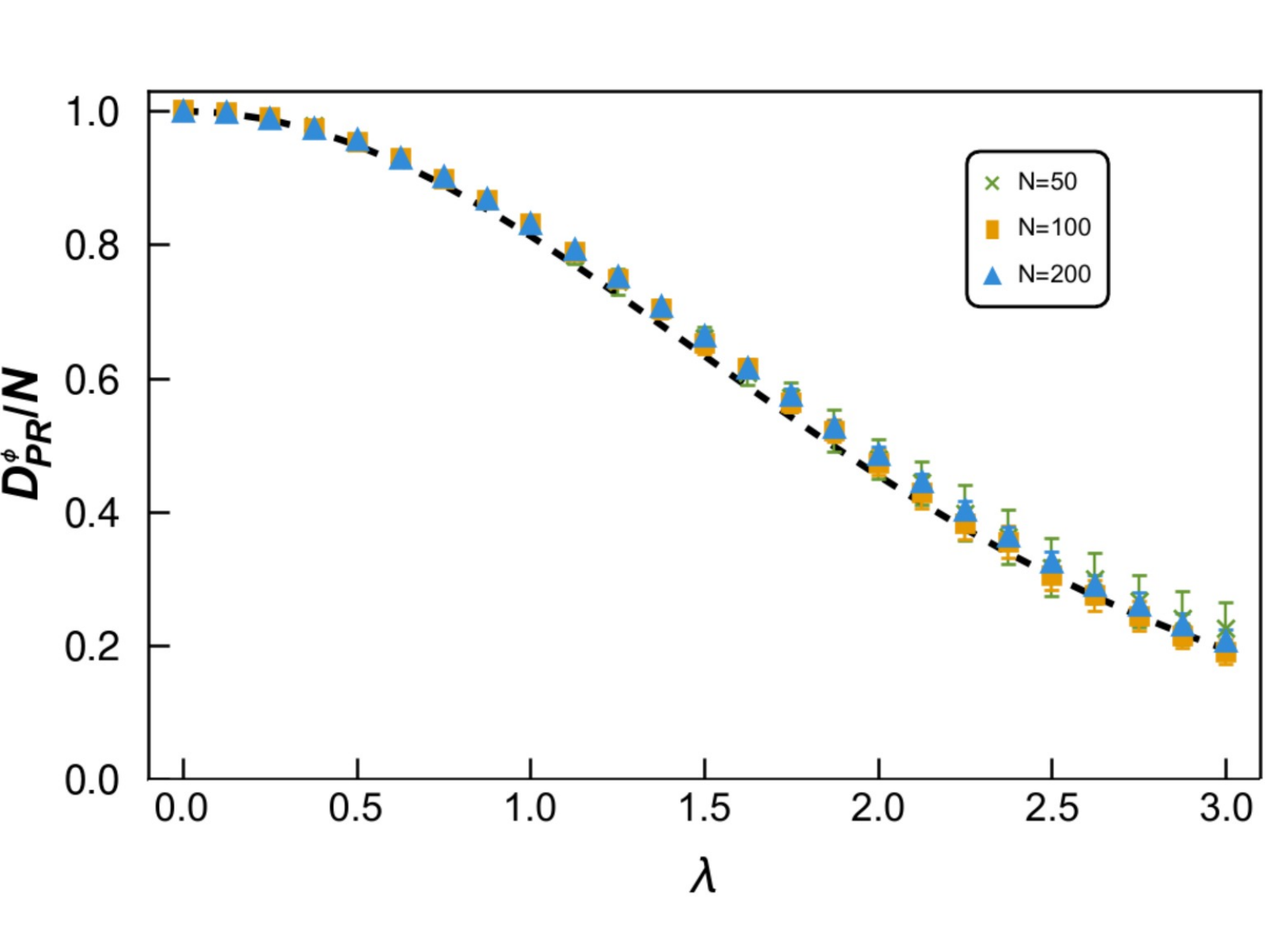}}\par\vspace{1mm}
    \subfigure{\includegraphics[width=0.32\textwidth]{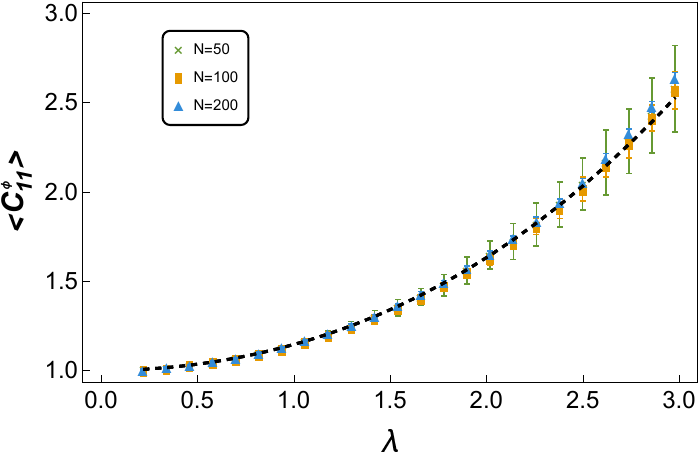}}\hfill
    \subfigure{\includegraphics[width=0.31\textwidth]{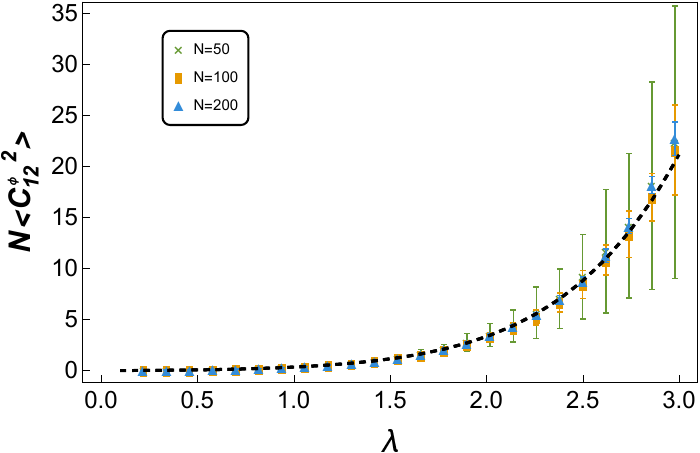}}\hfill
    \subfigure{\includegraphics[width=0.31\textwidth]{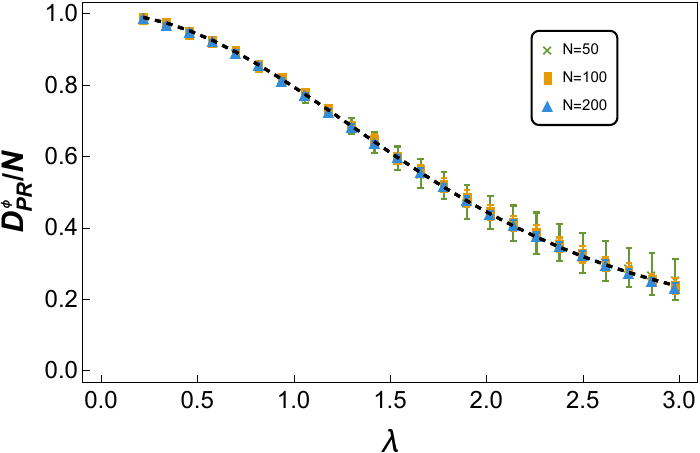}}\par\vspace{1mm}
    \subfigure{\includegraphics[width=0.31\textwidth]{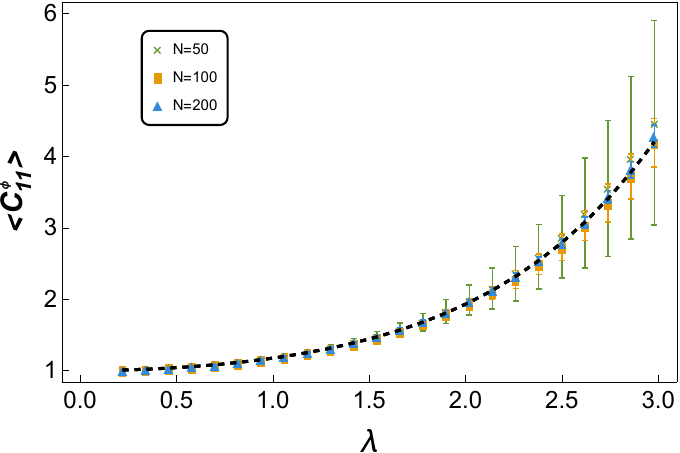}}\hfill
    \subfigure{\includegraphics[width=0.31\textwidth]{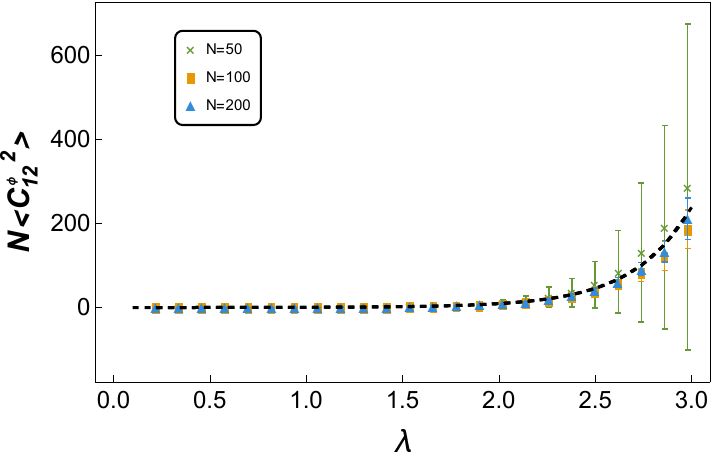}}\hfill
    \subfigure{\includegraphics[width=0.31\textwidth]{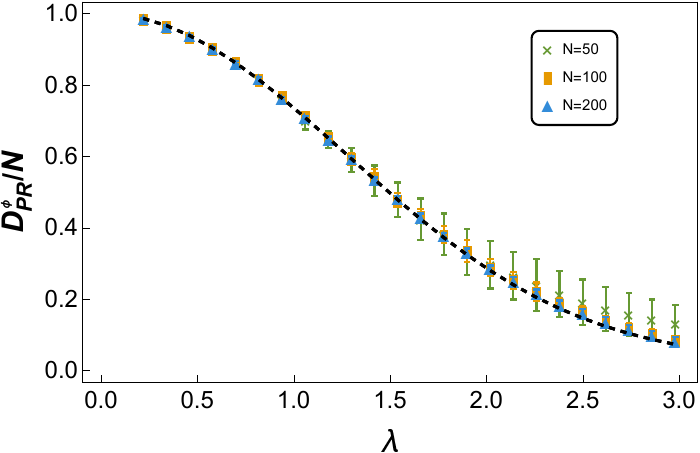}}
    \caption{Comparison of analytical and numerical results for the input correlation. Left column: average values of the diagonal correlations. Central column: average off-diagonal correlators. Right column: participation ratio (participation dimension divided by network size). In the upper row we use the transfer function of Eq.(\ref{eq:powerf}) with $\alpha=0.35$, $p=1/2$ and $D=1$; in the center we use the transfer function of Eq.(~\ref{eq:padep0}) with $\beta=2$ and $D=1$, and in the lower row we use the non saturating activation function of Eq.~(\ref{eq:padep1o2}). Network sizes $N=50, 100, 200$. The error bars represent the standard deviation over 5 simulations with different realizations of the coupling matrix $W$ and noise $\xi$.}
    \label{fig:cinp0}
\end{figure}

The comparison between the analytical results and  numerical simulations for the input correlations are shown in Figs.~\ref{fig:cinp0}. As in the case of output correlations we found an excellent agreement even for not very  large network sizes.

\clearpage
\section{Network simulations}\label{app:numerics}

The differential equations that control the network dynamics in the quenched limit, Eq. (\ref{eq:eomeq}), are simulated using a first order Euler method. As we intend to be in the limit $\tau \rightarrow 0$ at each time step we choose a vector of random Gaussian variables $\xi_i$ $(1=1,...N)$ with zero mean and variance $D$ and iterate
\begin{equation}
 \phi_i(t+\delta)=\phi_i(t)(1-\epsilon)+\left( \sum_{j=1}^N W_{ij} f(\phi_j(t)) + \xi_i\right)\epsilon
\end{equation}
until the mean quadratic difference $d \equiv \frac{1}{N}\sum_{i=1}^N|\phi_i(t+\delta)-\phi_i(t)|^2$ is smaller than $10^{-4}$. We use $\delta=0.01$, $\epsilon=0.1$. This procedure is repeated $n_t=1000$ times to obtain the covariance matrices:
\begin{eqnarray}
 C^f_{ij} &=& <<f(\phi_i) f(\phi_j)>>-  <<f(\phi_i)>> <<f(\phi_j)>> \\
 C^{\phi}_{ij} &=& <<\phi_i \phi_j>>-  <<\phi_i>> <<\phi_j>>
\end{eqnarray}
where the brackets denote the trial average:
$<<x>>= \frac{1}{n_t} \sum_{l=1}^{n_t} x(l)$.
As finite sampling generates bias in the statistics, specifically in the second order terms, we implemented the corrections proposed in \cite{doi:10.1073/pnas.1818972116} (Supplementary Material).

The results shown in Section \ref{subsec:comparison} for the opposite limit ($\tau_{noise} \ll \tau$) were obtained by simulating Eq.(\ref{eq:eom11}) using the Euler-Maruyama method, i.e.
\begin{equation}
 \phi_i(t+\delta)=\phi_i(t)\left(1-\frac{\delta}{\tau}\right)+\left( \sum_{j=1}^N W_{ij} f(\phi_j(t))\right)\frac{\delta}{\tau} + \frac{\sqrt{\delta}}{{\tau}}\xi_i
\end{equation}
where $\xi_i$ is a vector of random variables with Gaussian distribution (zero mean and variance $D$) that are extracted independently at each time step. The results of the simulations were used to evaluate the time lagged correlation functions in the same way as in \cite{shen2025covariance}.

\begin{figure}[h]
    \centering
     \includegraphics[width=1\textwidth]{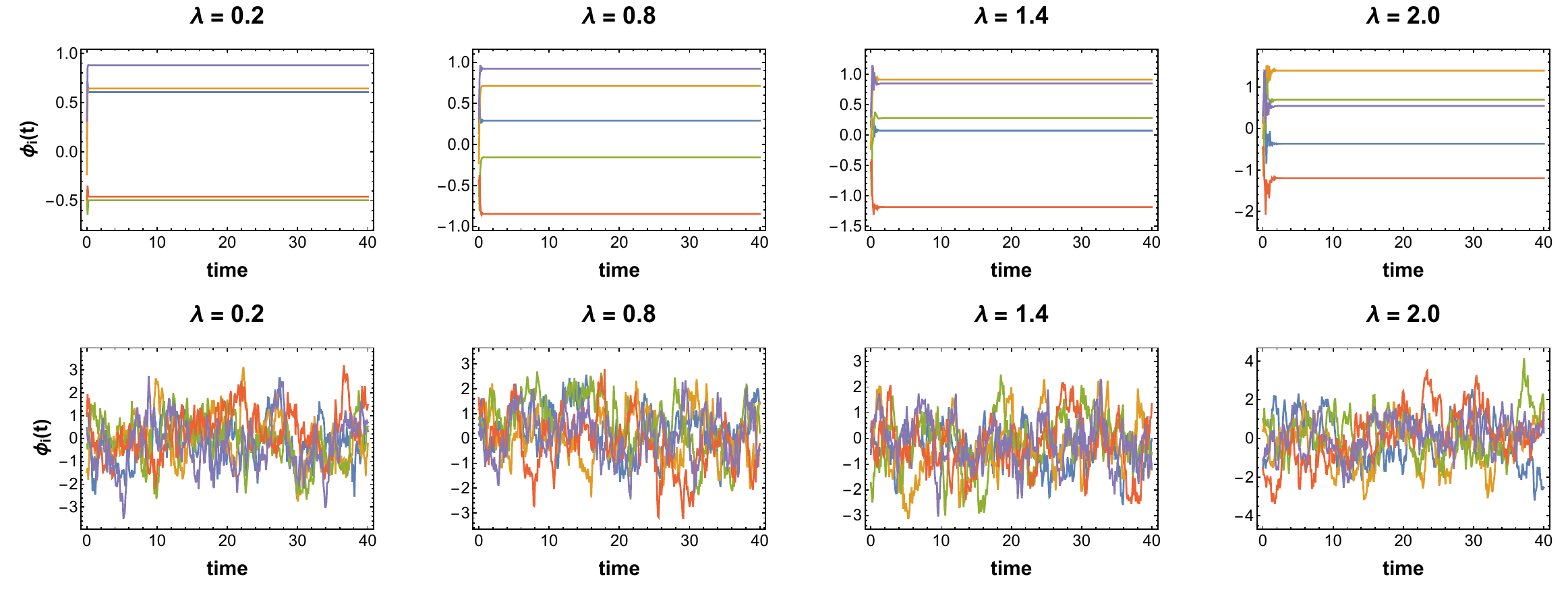}
    \caption{Comparison of dynamical behavior for the quenched (upper panels)  and annealed dynamics (lower panels). Sample traces of 5 neurons for different values of $\lambda$. Parameters as in Fig. \ref{fig:coutp0} with $N=100$. For the quenched dynamics, $\delta=0.01$ and $\epsilon=0.1$, so that $\tau=0.1$. Thus the time $t=2$ shown in the upper panels corresponds to $20\tau$.}
    \label{fig:traces}
\end{figure}

In Fig. \ref{fig:traces} we show representative traces for the two dynamics. In the upper panels we display the convergence of the quenched dynamics (Eq. (\ref{eq:eomeq})) to the fixed point determined by the values of the noise variable. In the lower panel we show the very fast changes induced by the annealed dynamics of Eq. (\ref{eq:eom11}).

\newpage
\section{Derivation of the effective action at large $N$ at order $J^2$}\label{app:deriv1}

In this Appendix for completeness we present the details of the derivation of the effective action (\ref{eq:Seff2}).

We will consider an expansion for the exponential that resums the sources at each order. For a single neuron index $j$ we have
\begin{align}
\label{eq:appcumulant_expand}
	\langle e^{-\tfrac{i}{2} \tr(\eta g)+\tr \left(J_j (g+xx) \right)} \rangle_G
	&= \Big\langle e^{-\tfrac{i}{2} \tr(\eta g)}\Big(1+ \tr \!\left(J_j (g+xx) \right) 
	+ \tfrac{1}{2!} \big(\tr \!\left[J_j (g+xx) \right]\big)^2+ \ldots\Big) \Big\rangle_G 
	\notag \\[1ex]
	&\hspace{-3cm} = \langle e^{-\tfrac{i}{2} \tr(\eta g)}\rangle_G \,
	\left(1+\frac{\Big\langle e^{-\tfrac{i}{2} \tr(\eta g)} \tr \!\left(J_j (g+xx) \right)
		+ \tfrac{1}{2!} e^{-\tfrac{i}{2} \tr(\eta g)} \big(\tr \!\left[J_j (g+xx) \right]\big)^2+\ldots 
		\Big\rangle_G}
	{\langle e^{-\tfrac{i}{2} \tr(\eta g)}\rangle_G}\right) \notag \\[1ex]
	&\hspace{-3cm}= \langle e^{-\tfrac{i}{2} \tr(\eta g)}\rangle_G\,
	\textrm{exp}\!\left[ 
	\frac{\langle e^{-\tfrac{i}{2} \tr(\eta g)} \, \tr \!\left(J_j (g+xx) \right)\rangle_G}
	{\langle e^{-\tfrac{i}{2} \tr(\eta g)}\rangle_G}
	\right] \\[1ex]
	&\hspace{-3cm} \times \textrm{exp}\!\left[
\tfrac{1}{2} \,
\frac{\langle e^{-\tfrac{i}{2} \tr(\eta g)} \big(\tr \!\left[J_j (g+xx) \right]\big)^2\rangle_G}
{\langle e^{-\tfrac{i}{2} \tr(\eta g)}\rangle_G}-\tfrac{1}{2}\frac{ \big(\langle e^{-\tfrac{i}{2} \tr(\eta g)}  \tr \!\left(J_j (g+xx) \right) \rangle_G\big)^2}{(\langle e^{-\tfrac{i}{2} \tr(\eta g)}\rangle_G)^2}
+ \ldots\right]. \notag
\end{align}

Similarly, the expectation value for the product over $j$ indices becomes
\begin{align}
\label{eq:appcumulant_expand2}
\langle \prod_j e^{-\tfrac{i}{2} \tr(\eta g)+\tr \left(J_j (g+xx) \right)} \rangle_G
&=\langle e^{-\tfrac{i}{2} \tr(\eta g)}\rangle^N_G\,
   \textrm{exp}\!\left[\sum_j 
     \frac{\langle e^{-\tfrac{i}{2} \tr(\eta g)} \, \tr \!\left(J_j (g+xx) \right)\rangle_G}
          {\langle e^{-\tfrac{i}{2} \tr(\eta g)}\rangle_G}
   \right] \\[1ex]
&\hspace{-4cm} \times \textrm{exp}\!\left[
   \tfrac{1}{2}\sum_j \,
   \frac{\langle e^{-\tfrac{i}{2} \tr(\eta g)} \big(\tr \!\left[J_j (g+xx) \right]\big)^2\rangle_G}
        {\langle e^{-\tfrac{i}{2} \tr(\eta g)}\rangle_G}-\tfrac{1}{2}\sum_j\frac{ \big(\langle e^{-\tfrac{i}{2} \tr(\eta g)}  \tr \!\left(J_j (g+xx) \right) \rangle_G\big)^2}{(\langle e^{-\tfrac{i}{2} \tr(\eta g)}\rangle_G)^2}
   + \ldots\right]\,. \notag
\end{align}
In the second line of this expression, we used the fact that contributions from different
indices j factorize in the expectation value and hence cancel out between the two terms. By
generalizing this procedure to higher orders we can compute the probability distribution for
all neural correlators.

As discussed in the main text, we focus for concreteness on calculating the partition function including second order contributions from the sources. Terms at higher orders an be obtained following the same procedure. We write this as
\be\label{eq:appZnrfinal}
Z_{N_r}(J) = \int  {\mathcal D}\eta\,  {\mathcal D}\rho\, e^{-N S_{eff}}\,.
\ee

Inserting \eqref{eq:appcumulant_expand2} into \eqref{eq:Znrnoexpand} and collecting all factors in the exponential,
we obtain, up to $\mathcal{O}(J^2)$, 
\begin{align}
S_{\mathrm{eff}}
&=\frac{1}{2}\log\det(1+iM\eta)
-\frac{i}{2}\tr(\eta\rho)
-\log\Big\langle e^{-\frac{i}{2}\tr(\eta g(x))}\Big\rangle_G
\nonumber\\
&\quad
-\frac{1}{N}\sum_{j}
\frac{\Big\langle e^{-\frac{i}{2}\tr(\eta g(x))}\,
\tr\!\big(J_j (g(x)+xx)\big)\Big\rangle_G}
{\Big\langle e^{-\frac{i}{2}\tr(\eta g(x))}\Big\rangle_G}
\nonumber\\
&\quad
-\frac{1}{2N}\sum_{j}
\frac{\Big\langle e^{-\frac{i}{2}\tr(\eta g(x))}\,
\big(\tr\!\big[J_j (g(x)+xx)\big]\big)^2\Big\rangle_G}
{\Big\langle e^{-\frac{i}{2}\tr(\eta g(x))}\Big\rangle_G}
\nonumber\\
&\quad
+\frac{1}{2N}\sum_{j}\frac{\Big(\Big\langle e^{-\frac{i}{2}\tr(\eta g(x))}\,
\tr\!\big(J_j (g(x)+xx)\big)\Big\rangle_G\Big)^2}
{\Big(\Big\langle e^{-\frac{i}{2}\tr(\eta g(x))}\Big\rangle_G\Big)^2}+\ldots
\label{eq:Seff_J2_ratio}
\end{align}

The cumulant expansion of the first expectation value gives
\begin{align}
-\log\Big\langle e^{-\frac{i}{2}\tr(\eta g)}\Big\rangle_G
&=\frac{i}{2}\tr\!\big(\eta\langle g\rangle_G\big)
+\frac{1}{8}\Big(\big\langle(\tr(\eta g))^2\big\rangle_G
-\big\langle\tr(\eta g)\big\rangle_G^{\,2}\Big)
+\mathcal{O}(\eta^3).
\label{eq:log_expand_1}
\end{align}
Note that \eqref{eq:appcumulant_expand2} is a cumulant expansion in the sources $J_j$ (keeping the $\eta$-dependent weight exact),
whereas \eqref{eq:log_expand_1} follows from a separate cumulant expansion in $\eta$. Using the symmetry of $\eta$ and $g$,
\begin{equation}
\tr(\eta g)=\sum_{a}\eta_{aa}g_{aa}+2\sum_{a<b}\eta_{ab}g_{ab},
\label{eq:trace_symm_expand}
\end{equation}
and expanding the $\eta$-dependence:
\begin{align}
-\log\Big\langle e^{-\frac{i}{2}\tr(\eta g)}\Big\rangle_G
&\approx\frac{i}{2}\tr\!\big(\eta\langle g\rangle_G\big)
+\frac{1}{2}\sum_{a<b}\eta_{ab}^2\,\langle g_{ab}^2\rangle_G.
\label{eq:log_expand_2}
\end{align}

For the linear source contribution we expand the ratio in \eqref{eq:Seff_J2_ratio} to first
order in $\eta$:
\begin{align}
\frac{\Big\langle e^{-\frac{i}{2}\tr(\eta g)}\,\tr\!\big(J_j (g+xx)\big)\Big\rangle_G}
{\Big\langle e^{-\frac{i}{2}\tr(\eta g)}\Big\rangle_G}
&=\Big\langle \tr\!\big(J_j (g+xx)\big)\Big\rangle_G
-\frac{i}{2}\Big\langle \tr(\eta g)\,\tr\!\big(J_j (g+xx)\big)\Big\rangle_G
\nonumber\\
&\quad
+\frac{i}{2}\Big\langle \tr(\eta g)\Big\rangle_G
\Big\langle \tr\!\big(J_j (g+xx)\big)\Big\rangle_G
+\mathcal{O}(\eta^2 J),
\label{eq:ratio_expand}
\end{align}
To obtain an explicit expression in replica indices, we write
$\tr\!\big(J_j(g+xx)\big)=\sum_{ab}J_j^{ab}\big(g_{ab}+x^a x^b\big)$ and use
$\langle x^a x^b\rangle_G=G_{ab}$. We then separate diagonal and independent off-diagonal components
in replica space and keep only the contributions that survive at the
leading orders under the large-$N$ scaling stated above. In particular, in the $\mathcal{O}(\eta J)$ term we
retain the off-diagonal sector $a\neq b$ and the leading pairwise contractions at large $N$, which
yields
\begin{align}
\frac{\Big\langle e^{-\frac{i}{2}\tr(\eta g)}\,\tr\!\big(J_j (g+xx)\big)\Big\rangle_G}
{\Big\langle e^{-\frac{i}{2}\tr(\eta g)}\Big\rangle_G}
&=\sum_{a} J_j^{aa}\Big(G_{aa}+\langle g_{aa}\rangle_G\Big)
+\sum_{a<b} J_j^{ab}\Big(G_{ab}+\langle g_{ab}\rangle_G\Big)
\nonumber\\
&\quad
-i\sum_{a<b}\eta_{ab}\,J_j^{ab}\,
\Big\langle g_{ab}\big(g_{ab}+x^a x^b\big)\Big\rangle_G
+O(\eta^2 J).
\label{eq:ratio_expand_final}
\end{align}
For the quadratic source contribution we only need the leading term at $\eta=0$, since any
$\eta$-dependence would produce subleading corrections of order $O(\eta J^2)$ under the same scaling:
\begin{align}
&\frac{\Big\langle e^{-\frac{i}{2}\tr(\eta g)}\,
\big(\tr\!\big[J_j (g+xx)\big]\big)^2\Big\rangle_G}
{\Big\langle e^{-\frac{i}{2}\tr(\eta g)}\Big\rangle_G}
-\frac{\Big(\Big\langle e^{-\frac{i}{2}\tr(\eta g)}\,
\tr\!\big(J_j (g+xx)\big)\Big\rangle_G\Big)^2}
{\Big(\Big\langle e^{-\frac{i}{2}\tr(\eta g)}\Big\rangle_G\Big)^2}
\nonumber\\
&\qquad=
\sum_{a}(J_j^{aa})^2
\Big(\big\langle (g_{aa}+x^a x^a)^2\big\rangle_G-\big\langle g_{aa}+x^a x^a\big\rangle_G^2\Big)
+\sum_{a<b}(J_j^{ab})^2\,\big\langle (g_{ab}+x^a x^b)^2\big\rangle_G
+O(\eta J^2),
\label{eq:ratio_expand_quadratic_final}
\end{align}

We have checked  self-consistently that the diagonal components of $\rho_{ab}$ and $G_{ab}$ are order $N^0$, while the off-diagonal components of $\rho_{ab}$ and $G_{ab}$, as well as all the variables $\eta_{ab}$ are order $1/N$. With this $N$-scaling, the leading terms at large $N$ are
\begin{align}\label{eq:appSeff2}
S_{eff}&\approx \frac{1}{2} \log \det(1+i M \eta)-\frac{i}{2}\tr( \eta \rho)+ \frac{i}{2} \tr\left(\eta \langle g \rangle_G\right)+\frac{1}{2} \sum_{a<b}\eta_{ab}^2\langle g_{ab}^2 \rangle_G  \\
&-\frac{1}{N}\sum_j \left[ \sum_a J_j^{aa} \left(G_{aa}+ \langle g_{aa}\rangle_G\right) + \sum_{a<b} J_j^{ab}\left( G_{ab}+ \langle g_{ab}\rangle_G - i \eta_{ab} \langle g_{ab}(g_{ab}+x^a x^b )\rangle_G\right)\right]\notag\\
&-\frac{1}{2N}\sum_j \left[ \sum_a  (J_j^{aa})^2 \left(\langle(g_{aa}+x^a x^a)^2 \rangle_G-\langle g_{aa}+x^a x^a \rangle_G^2 \right) +\sum_{a<b} (J_j^{ab})^2 \langle (g_{ab}+ x^a x^b)^2 \rangle_G \right]\notag\,,
\end{align}
which reproduces (\ref{eq:Seff2}).

\bibliography{RNN}{}
\bibliographystyle{utphys}

\end{document}